\documentclass[aps, prd, superscriptaddress, nofootinbib, preprintnumbers]{revtex4-2}
\usepackage[colorlinks=true]{hyperref}
\usepackage{xcolor}
\hypersetup{colorlinks=true, citecolor=blue, urlcolor=blue, linkcolor=blue}
\usepackage{amsmath,amssymb}
\usepackage[final]{graphicx}
\usepackage{subcaption}
\usepackage[belowskip=0.5pt,aboveskip=0.5pt]{caption}
\usepackage{float}
\usepackage{adjustbox}
\usepackage{cancel,soul,ulem}

\begin{document}

\title{Gluon polarization contribution to the spin alignment of vector mesons from holography }

\author{Hiwa A. Ahmed}
\email{hiwa.a.ahmed@mails.ucas.ac.cn}
\affiliation{School of Nuclear Science and Technology, University of Chinese Academy of Sciences, Beijing, 100049, P.R. China}
\affiliation{Department of Physics, College of Science, Charmo University, 46023 Chamchamal/Sulaimani, Kurdistan region-lraq}

\author{Yidian Chen}
\email{chenyidian@hznu.edu.cn}
\affiliation{School of Physics, Hangzhou Normal University, Hangzhou, 311121, P.R. China}

\author{Mei Huang}
\email{huangmei@ucas.ac.cn}
\affiliation{School of Nuclear Science and Technology, University of Chinese Academy of Sciences, Beijing, 100049, P.R. China}

\date{\today}

\begin{abstract}

We investigate the behaviour of vector mesons $\rho$, $\phi$, and $J/\Psi$ in both non-rotating and rotating thermal media using the soft-wall holographic QCD model with four flavours. By incorporating anisotropic backgrounds derived from the Einstein-Maxwell-dilaton action, we incorporate rotational effects via a $U(1)$ gauge field, and the induced polarization of gluons is described by a rotation dependent dilation field. Spectral function analysis reveals that $\rho$ and $\phi$ mesons exhibit broad peaks at lower temperatures, indicating their presence in the medium, while these peaks disappear at higher temperatures. Rotation delays this melting process, increasing the dissociation temperature. In contrast, the $J/\Psi$ meson, owing to its heavy charm quark content, demonstrating its resilience to thermal effects. We further explore the global spin alignment of these mesons in the event plane frame. For the $\phi$ meson, the averaged $\rho_{00}$ over the full range of azimuthal angle shows weak temperature dependence at low transverse momentum ($p_T$) but significant suppression at high $p_T$, aligning with experimental observations. Rotation enhances $\rho_{00}$ at high $p_T$, a phenomenon attributed to angular momentum transfer via spin-orbit coupling. The $J/\Psi$ meson, however, displays insensitivity to temperature and rotation up to $p_T=5$ GeV, with a very small suppression observed at higher $p_T$, likely due to its heavy quark nature. Although $\rho$ meson spin alignment is not yet experimentally measured, it exhibits behaviour qualitatively similar to the $\phi$ meson, with thermal fluctuations dampening alignment and rotation enhancing it.

\end{abstract}


\maketitle

\section{Introduction}

The study of vector mesons in high-energy heavy ion collisions provides a unique window into the properties of the quark-gluon plasma (QGP), a state of matter where quarks and gluons are deconfined. Among the various observables, the global spin alignment of vector mesons has emerged as a powerful probe to investigate the interplay between the collective dynamics of QGP and the intrinsic spin properties of hadrons. Spin alignment, characterized by the parameter $\rho_{00}$, measures the probability of finding a vector meson in a spin state aligned with a chosen quantization axis. Deviations from the statistical value of $1/3$ indicate the presence of non-trivial spin-dependent interactions in the medium, offering insights into the mechanisms of spin-orbit coupling, thermal fluctuations, and rotational effects in the QGP.

In a non-central heavy ion collision, a substantial orbital angular momentum is generated, which can induce spin polarization in hyperons and spin alignment in vector mesons. This phenomenon was initially predicted in Refs. \cite{Liang:2004ph,Liang:2004xn}, and has been experimentally confirmed by collaborations such as STAR \cite{STAR:2017ckg,STAR:2018gyt,STAR:2019erd,STAR:2022fan} and ALICE \cite{ALICE:2019aid,ALICE:2020iev,ALICE:2022dyy}. Recent STAR measurements show that the $\rho_{00}$ of the $\phi$ meson deviates from the statistical value of $1/3$, with the deviation increasing as collision energy decreases \cite{STAR:2022fan}. However, the $\rho_{00}$ of $k^{*}$ displays non-monotonic behaviour in $20\%-60\%$ centrality collisions, while ALICE reports a significant spin alignment of $J/\Psi$ in $5.02$ TeV Pb-Pb collisions \cite{ALICE:2022dyy}. 
Theoretical efforts to explain spin alignment have explored various mechanisms. Global rotation and magnetic fields, generated in non-central collisions, are often considered primary drivers of spin polarization and alignment. While global rotation leads to $\rho_{00} \le 1/3$ \cite{Wei:2023pdf}, magnetic fields induce $\rho_{00} \ge 1/3$ \cite{Sheng:2022ssp}. However, the strengths of these fields in heavy ion collisions are believed to be too weak to produce the observed deviations in $\rho_{00}$ \cite{Yang:2017sdk,Xia:2020tyd}.

Recent theoretical developments propose alternative mechanisms, such as a vector field in strong interaction (called the $\phi$ field) coupled to the strange and antistrange quark plays an important role in the $\rho_{00}$ of $\phi$ meson \cite{Sheng:2019kmk,Sheng:2022wsy,Sheng:2023urn}, self-energy modifications of quarks \cite{Fang:2023bbw}, local axial charge current fluctuations \cite{Muller:2021hpe}, helicity polarization \cite{Gao:2021rom}, local vorticity and anisotropic expansion of the fireball \cite{Xia:2020tyd}, plasma fields \cite{Kumar:2023ghs}, light-front quark dynamics \cite{Fu:2023qht} and local spin density fluctuations\cite{Xu:2024kdh}. Additionally, the relationship between tensor polarization and quark spin correlations has been explored in \cite{Lv:2024uev}. 

A particularly puzzling observation is the spin alignment of $\phi$ mesons in most central collisions ($0-20\%$) at STAR \cite{STAR:2022fan}, where $\rho_{00}$ is smaller than $1/3$ at high energies but larger than $1/3$ at low energies. This behaviour contradicts traditional expectations, as central collisions are associated with weaker magnetic fields, rotation, and baryon chemical potential, suggesting that $\rho_{00}$ should approach $1/3$. This discrepancy raises the question of whether spin alignment arises from intrinsic QCD dynamics rather than external factors such as rotation or magnetic fields.

Besides spin alignment, the study of QCD matter and QCD phase transitions also attracted significant attention.  From effective QCD models \cite{Chen:2015hfc,Jiang:2016wvv,Ebihara:2016fwa,Chernodub:2016kxh,Chernodub:2017ref,Wang:2018sur,Sun:2021hxo, Xu:2022hql,Sun:2023kuu} as well as holographic QCD  models \cite{Chen:2020ath,Braga:2023qej,Li:2023mpv, Ambrus:2023bid,Zhao:2022uxc,Golubtsova:2022ldm, Chen:2022mhf,Chen:2022smf,Yadav:2022qcl,Braga:2022yfe, Cartwright:2021qpp,Golubtsova:2021agl}, it has been found that the rotation suppresses the critical temperature of chiral phase transition. Opposite results from lattice QCD (LQCD) calculations \cite{Braguta:2021jgn,Braguta:2022str,Yang:2023vsw} showed that both critical temperatures of the deconfinement and chiral symmetry restoration phase transitions increase with angular velocity.  This puzzle has attracted much attention, and the polarization of gluon degrees of freedom under rotation \cite{Sun:2024anu,Chen:2024jet} offered a reasonable understanding of this problem. 

A comprehensive understanding of the vector meson spin alignment and QCD phase transitions under rotation requires a nonperturbative framework capable of describing the strongly coupled nature of the QGP. Holographic QCD, based on the gauge/gravity duality, provides such a framework by mapping the dynamics of strongly coupled gauge theories to a higher-dimensional gravitational theory \cite{Maldacena:1997re,Witten:1998qj}. In this work, we investigate the possibility that spin alignment generated by the intrinsic dynamics of QCD such as gluon polarization and what is the effect of temperature and rotation on the spin alignment using the bottom-up soft-wall holographic QCD model with four flavors.

The soft-wall model, which incorporates a dilaton field to generate confinement\cite{Karch:2006pv}, provides a nonperturbative framework for analyzing the spectral functions \cite{Son:2002sd} and spin alignment of vector mesons \cite{Sheng:2024kgg,Zhao:2024ipr}. To incorporate rotational effects, anisotropic backgrounds derived from the Einstein-Maxwell-dilaton (EMD) action (which is known as the dynamical holographic QCD (DHQCD)) have been considered, where rotation is modelled through a $U(1)$ gauge field and a dilaton field \cite{Chen:2022mhf,Chen:2024jet}. This approach allows us to systematically explore the influence of temperature, chemical potential, and angular velocity on the spin alignment of vector mesons. In order to study the vector mesons ($\rho$, $\phi$, and $J/\Psi$), we need to consider the four-flavor soft-wall model \cite{Chen:2021wzj,Ahmed:2023zkk}. 

Our study begins with an analysis of the deconfinement/confinement phase transition in the presence of rotation. We demonstrate that rotation increases the critical temperature of the phase transition, which is consistent with LQCD results, and identify the dominant role of the dilaton field in this enhancement. We then examine the spectral functions of $\rho$, $\phi$, and $J/\Psi$ mesons, which provide information about their dissociation temperatures and the modifications induced by the medium. The core of our investigation focuses on the global spin alignment of vector mesons in the event plane frame. For the $\phi$ and $J/\Psi$, we compare our results with the experimental data and then show the explicit dependence of the global spin alignment on the temperature and angular velocity. Finally, we predict the $\rho_{00}$ of the $\rho$ at finite temperature and angular velocity.

The paper is organized as follows. Firstly, we will briefly introduce the DHQCD model in an anisotropic background and introduce the four-flavor soft-wall model in this background in section \ref{sectionII}. Then, at the end of this section, we review the dilepton production through a vector meson decay and provide the relation between the spin alignment and the spectral function. In section \ref{sectionIII}, we will study the effect of rotation on the deconfinement/confinement phase transition and melting of the vector meson in a thermal and rotating medium. Towards the end of section \ref{sectionIII}, we will reach the primary goal of our work, which is the global spin alignment of the $\rho$, $\phi$, and $J/\Psi$ mesons.
 Finally, we will conclude our work in section \ref{sectionIIII}. 

\section{Model Set up}
\label{sectionII}

The holographic QCD approach, rooted in the Anti-de Sitter/conformal field theory (AdS/CFT) correspondence, has emerged as a significant nonperturbative technique for exploring the characteristics of QCD and its phase transitions. Building upon these foundational studies, the DHQCD has advanced our understanding of the dynamical behaviour of QCD by effectively capturing phenomena such as confinement/deconfinement transitions and chiral phase transitions. This paper aims to explore the influence of rotation on the spectral functions and spin alignment of vector mesons. To achieve this, it is essential to consider a rotational background geometry, for which DHQCD serves as a suitable framework. In this section, we will review the DHQCD model, including the aspect of rotation, followed by examining the formalism of vector mesons within the four-flavor soft-wall model. Lastly, we will outline the general methodology for calculating the spectral function and the spin density matrix using the holographic QCD model.

\subsection{The Einstein-Maxwell-dilaton system}

In this section, the formalism presented in Ref. \cite{Chen:2024jet} is examined with regard to the study of rotating systems. The EMD action within the string frame can be expressed as
\begin{equation}
S^s=\frac{1}{16 \pi G_5} \int d^5 x \sqrt{-g^s} e^{-2 \Phi}\left[R^s+4 \partial_M \Phi \partial^M \Phi-V^s(\Phi)-\frac{h(\Phi)}{4} e^{\frac{4 \Phi}{3}} F_{M N} F^{M N}\right],
\end{equation}
where the notation lowercase $s$ refers to the string frame, and $G_5$ represents the five-dimensional Newtonian constant. The fields $\Phi$ and $V_s(\Phi)$ signify the dilaton field and its corresponding potential, respectively. The strength tensor $F^{M N}$, representing the gauge field $A_M$ dual to the baryon number current, is also introduced. The coupling between the dilaton field and the gauge field is articulated by the function $h(\Phi)$. For convenience, the equations of motion and thermodynamic considerations are reformulated in the Einstein frame, yielding:
\begin{equation}
S^e=\frac{1}{16 \pi G_5} \int d^5 x \sqrt{-g^e}\left[R^e-\frac{1}{2} \partial_M \Phi \partial^M \Phi-V^e(\Phi)-\frac{h(\Phi)}{4} F_{M N} F^{M N}\right],
\end{equation}
where the lowercase $e$ is assigned for the Einstein frame. In this context, it is observed that the metric and potential are related by the equations $g_{M N}^s=e^{\frac{4 \Phi}{3}} g_{M N}^e$ and $V^e=e^{\frac{4 \Phi}{3}} V^s$. Within Einstein's framework, the metric can take the form
\begin{equation}
\left.d s^2=\frac{L^2 e^{2 A_e(z)}}{z^2}\left[-f(z) d t^2+\frac{d z^2}{f(z)}+d \vec{x}^2\right)\right],
\end{equation}
where the warp factors $A_e(z)$ and $A_s(z)$ satisfy the relation $A_e(z)=A_s(z)-\frac{2}{3} \Phi(z)$. For simplification, the AdS radius is set to unity, $L=1$.

To investigate a rotating system consistent with lattice QCD, an anisotropic background was proposed in Refs. \cite{Chen:2022mhf,Chen:2024jet}. In their formalism, the chiral condensate shows minimal dependence on the radial coordinate in proximity to the center of rotation. Thus, it is justifiable to consider that the metric and dilaton fields are predominantly functions of the fifth-dimensional coordinate near the center. Furthermore, to account for the rotational effects, it is essential to incorporate an additional anisotropic function $B(z)$ into the metric. Thus, the background metric of the Einstein frame within the cylindrical coordinate framework is expressed as
\begin{equation}
d s^2=\frac{ e^{2 A_e(z)}}{z^2}\left[-f(z) d t^2+\frac{d z^2}{f(z)}+e^{B(z)} d r^2+r^2 e^{B(z)} d \theta^2+e^{-2 B(z)} d x_3^2\right],
\end{equation}
where $f(z)$ signifies the blackening factor, which approaches zero at the black hole horizon denoted by $z_h$. In alignment with the holographic principle, the temperature of the system can be derived from the Hawking temperature formula:
\begin{equation}
T=\frac{\left|f^{\prime}\left(z_h\right)\right|}{4 \pi}
\label{temp}
\end{equation}

The introduction of rotation dynamics into the system is facilitated by employing a non-zero polar angle component, denoted as $A_\theta$, within the gauge field configuration. In the vicinity of the center approximation, the gauge field ( $A_M $) is to be approximated as follows:
\begin{equation}
A_M = \left(A_t, 0, 0, A_\theta, 0\right), \quad A_\theta = \Omega r^2 ~,
\end{equation}
where ($ A_z = 0 $) represents the gauge fixing, and ($\Omega $) denotes the angular velocity from the external vortical field. Although the gauge field ($A_\theta $) exhibits dependence on the radial coordinates ($r$), the invariant strength ($F^2$) remains independent of ($r$). Consequently, the fifth-dimensional coordinate ($z$) solely determines the metric and dilaton fields. 

The equations of motion of the EMD system that can be used to find the dilaton field and the unknown functions in the metric are given by
\begin{subequations}
\begin{align}
-z^2 e^{-2 A_e} h(\Phi)\left(A_t^{\prime 2}+\frac{8}{3} e^{-2 B} \Omega^2\right)+3 f^{\prime}\left(A_e^{\prime}-\frac{1}{z}\right)+f^{\prime \prime}=0 ,\\
\frac{B^{\prime 2}}{2}+\frac{4 \Phi^{\prime 2}}{9}+\frac{2 A_e^{\prime}}{z}-A_e^{\prime 2}+A_e^{\prime \prime}=0 ,\\
\frac{4 \Omega^2 z^2 h(\Phi) e^{-2\left(A_e+B\right)}}{3 f}+B^{\prime}\left(3 A_e^{\prime}+\frac{f^{\prime}}{f}-\frac{3}{z}\right)+B^{\prime \prime}=0 ,\\
A_t^{\prime}\left(A_e^{\prime}+\frac{h^{\prime}}{h}-\frac{1}{z}\right)+A_t^{\prime \prime}=0 ,
\end{align}
\label{eomemd}
\end{subequations}
where the ($^\prime$) denotes the derivative with respect to $z$. There are six unknown functions in Eq. \eqref{eomemd}, to solve these coupled equations, we need at least to provide two of the unknown functions. Following the work of Ref. \cite{Chen:2022mhf, Chen:2024jet}, the dilaton field $\Phi(z)$ and $h(\Phi)$ are treated as an input function. It is well known that both the bosons and fermions feel the rotation in the medium. Since the dilaton field corresponds to the gluonic operator in the boundary, it should be sensitive to rotation \cite{Sun:2024anu}. To capture the rotation-dependent, the following simple form of the dilaton field is considered
\begin{equation}
\Phi=\left(\mu_G+\mu_{\Omega} \Omega^2\right)^2 z^2 \tanh \left(\mu_{G_2}^4 z^2 /\left(\mu_G+\mu_{\Omega} \Omega^2\right)^2\right),
\end{equation}
with three free parameters $\mu_G$, $\mu_{G_2}$, and $\mu_{\Omega}$. The coefficient $\mu_\Omega$ captures the effects of gluon polarization induced by rotation, reflecting the spin-orbit coupling contributions in the medium (see Refs. \cite{Sun:2024anu,WeiMingHua:2020eee}). In the ultraviolet (UV) region, the dilaton field has $\Phi(z\to 0) \to \mu_{G^2}^{4} z^4$ for zero angular velocity, which is dual to the gauge-invariant gluon condensation. In the infrared (IR) region, the dilaton field has $\Phi(z\to \infty) \to \mu_{G}^{2} z^2$, which is a necessary condition for the mesons spectra to satisfy the linear Regge trajectory. Finally, the coupling between the dilaton field and the gauge field $h(\Phi)$ is chosen to be \cite{Chen:2024ckb}
\begin{equation}
    h(\Phi)=e^{- \Phi - A_e}.
\end{equation}
The form of $h(\Phi)$ is designed to align with the linear Regge slope of meson trajectories.

The selection of appropriate boundary conditions is essential for solving the equations of motion outlined in Eq. \eqref{eomemd}. At the UV boundary, the background geometry must exhibit characteristics of asymptotically $AdS_5$ space-time, thus yielding 
\begin{equation}
f(0) = 1, \quad A_e(0) = 0, \quad B(0) = 0.
\end{equation}

Conversely, at the IR boundary, the variables $A_e$ and $B$ must adhere to the natural boundary conditions, while the blackening factor is constrained such that $f(z_h) = 0$. In accordance with the holographic principle, the gauge field $A_t(z)$ is required to fulfill the following boundary conditions:
\begin{equation}
A_t(0) = \mu, \quad A_t(z_h) = 0.
\end{equation}

\subsection{ Vector mesons in soft-wall holographic model}

After fixing the background of the system, we can study the vector mesons using the four-flavor holographic QCD. According to the holographic model, there is a correspondence between the 4D operators and corresponding 5D gauge fields \cite{Erlich:2005qh}. The operators and corresponding gauge fields that play a role in chiral dynamics are defined as follows:
\begin{equation}
J_{R/L \mu}^a=\bar{\psi}_{q R/L} \gamma_\mu t^a \psi_{q R/L} \to R_{\mu}^{a}/L_{\mu}^{a},
\end{equation}
where $J_{R/L \mu}^a$ are right/left-handed currents that are associated with the gauge fields $R_{\mu}^{a}$ and $L_{\mu}^{a}$. It is important to mention that $t^{a}$ with $a=1,2, ..., N_{f}^{2}-1$ are the generators of the $SU(N_{f})$ group. The general five-dimensional action for the vector field can be expressed as:
\begin{equation}
\begin{aligned}
S_{M} &=-\int d^{5} x \sqrt{-g} e^{-\phi} \operatorname{Tr}\left\{\frac{h(\phi)}{2 g_{5}^{2}} V^{M N} V_{M N} +\left(D^{M} H\right)^{\dagger}\left(D_{M} H\right)+ M_{5}^{2}|H|^{2}\right\},
\end{aligned}
\label{actionnew}
\end{equation}
where $D^{M} H = \partial_{M} H - i V_{M} H - i H V_{M} $ reflects the covariant derivative of the scalar field $H$, $M_{5}^{2}=(\Delta - p)(\Delta + p -4) =-3$ by taking the conformal dimension of the scalar field operator $\Delta=3$ and $p=0$. The coupling constant $g_{5}$ is defined as $g_{5} = 2 \pi $ for $N_{c}=3$  \cite{Erlich:2005qh}. The gauge field strength $V_{M N}$ for the diagonal vector mesons ($\rho$, $\phi$, and $J/\Psi$) is given by
\begin{equation}
V_{M N}=\partial_{M} V_{N}-\partial_{N} V_{M}, \\
\end{equation}
where the vector field is written in terms of the right- and left-handed gauge fields as $V_{M}=\frac{1}{2}(L_{M} + R_{M})$. The field $V_{M}$ can be expanded to $V_{M}^{a} t^{a}$, and the generators satisfy $Tr(t^{a} t^{b})=\frac{1}{2} \delta^{ab}$. The heavy scalar field $H$ is introduced into the action to explicitly break the $SU(4)_V$ symmetry of the diagonal vector fields to $SU(2)_V$ symmetry \cite{Chen:2021wzj,Ahmed:2023zkk}. The auxiliary field $H$ is a diagonal matrix. However, it only reflects the effect of the strange and charm quark mass, $H=\frac{1}{2}\operatorname{diag}\left[0, 0, h_{s}(z),h_{c}(z)\right]$. Then, $h_{c(s)}$ at the UV boundary should behave like $ h_{c(s)}(z\rightarrow 0)=m_{hc(s)} z $. In order to consider the two-point correlation function, we can expand the action up to the second order as the following
\begin{equation}
\begin{aligned}
S^{(2)}=&-\int d^{5}x \sqrt{-g} e^{-\phi(z)} \left\{\frac{h(\phi)}{4 g_{5}^{2} } g^{M P} g^{N Q}V^{a}_{M N} V^{b}_{P Q} + g^{M N} V_{M}^{a} V_{N}^{b} M_{H }^{a b}\right\}
\end{aligned}
\label{S2}
\end{equation}
where $g^{M N}$ is the metric of the anisotropic medium. The mass term in the action $M_{H}^{a b}$ is defined by
\begin{equation}
M_{H}^{a b}(z) = {\rm Tr}\left( \{H,t^{a}\} \{H,t^{b}\} \right),
\end{equation}
where $M_{H}^{a,b}$ is non-zero only for the cases $a=b=8$ and $a=b=15$. The equation of motion for the vector field is obtained from the action Eq. \eqref{S2},
\begin{equation}
 \partial_{M}( \frac{h(\phi) \sqrt{-g} e^{-\phi}}{2g_{5}^{2} } V^{M N}) - \sqrt{-g} e^{-\phi}   M_{H}^{a a}(z) V^{M}=0.
\label{emVm}
\end{equation}

By choosing the gauge fixing $V_z=0$ and the Fourier transformation for the remaining components,
\begin{equation}
V_\mu(x, z)=\int \frac{d^4 p}{(2 \pi)^4} e^{-i \omega t +i \mathbf{p} \cdot \mathbf{x}} V_\mu(p, z),
\end{equation}
the equation of motion becomes 
\begin{equation}
\begin{aligned}
& \partial_z\left[h(\phi)\frac{\sqrt{-g} e^{-\phi}}{2g_{5}^{2} } g^{z z} g^{\mu \nu} \partial_z V_\nu(p, z)\right]-  p_\alpha\left[\frac{\sqrt{-g} e^{-\phi}}{2g_{5}^{2} } g^{\alpha \beta} g^{\mu \nu}\left(p_\beta V_\nu(p, z)-p_\nu V_\beta(p, z)\right)\right]\\
& - \sqrt{-g} e^{-\phi} g^{\mu \nu}  M_{H}^{a a}(z) V_\nu(p, z) =0, \\
& g^{\mu \nu} p_\mu \partial_z V_\nu(p, z)=0,
\end{aligned}
\end{equation}
where the four-momentum $p_\mu$ is given as $p_\mu=(\omega, \mathbf{p})$. For simplicity, we work in the Cartesian coordinate, and the metric becomes
\begin{equation}
d s^2=\frac{ e^{2 A_e(z)}}{z^2}\left[-f(z) d t^2+\frac{d z^2}{f(z)}+e^{B(z)} (d x_1^2+ d x_2^2)+e^{-2 B(z)} d x_3^2\right].
\end{equation}

Within the above background, the equation of motion for the components of the gauge field is given by
\begin{equation}
\begin{aligned}
& V_t^{\prime \prime}+\left(-\frac{1}{z}-\phi^{\prime}+\frac{h^{\prime}}{h}+A_e^{\prime}\right) V_t^{\prime} -\frac{ 2 g_5^{2} e^{2 A_e}}{f h z^2}M_{H}^{a a}(z)  V_t  \\
& -\frac{e^{-B}}{f}\left(p_{x_1}\left(p_{x_1} V_t+\omega V_{x_1}\right)+p_{x_2}\left(p_{x_2} V_t+\omega V_{x_2}\right)+e^{3 B}p_{x_3}\left(p_{x_3} V_t+\omega V_{x_3}\right)\right)  =0 \\
& V_{x_1}^{\prime \prime}+\left(-\frac{1}{z}-\phi^{\prime}+\frac{h^{\prime}}{h}+A_e^{\prime}+\frac{f^{\prime}}{f}-B^{\prime}\right) V_{x_1}^{\prime} -\frac{ 2 g_5^{2} e^{2 A_e}}{f h z^2} M_{H}^{a a}(z) V_{x_1} \\  & +\frac{1}{f^2}\left(\omega \left(\omega V_{x_1}+p_{x_1} V_{t}\right)-f e^{-B} p_{x_2}\left(p_{x_2} V_{x_1}-p_{x_1} V_{x_2}\right)-f e^{2 B}p_{x_3}\left(p_{x_3} V_{x_1}-p_{x_1} V_{x_3}\right)\right)  =0 \\
& V_{x_2}^{\prime \prime}+\left(-\frac{1}{z}-\phi^{\prime}+\frac{h^{\prime}}{h}+A_e^{\prime}+\frac{f^{\prime}}{f}-B^{\prime}\right) V_{x_2}^{\prime} -\frac{ 2 g_5^{2} e^{2 A_e}}{f h z^2} M_{H}^{a a}(z) V_{x_2} \\ & +\frac{1}{f^2}\left(\omega \left(\omega V_{x_2}+p_{x_2} V_{t}\right)-f e^{-B} p_{x_1}\left(p_{x_1} V_{x_2}-p_{x_2} V_{x_1}\right)-f e^{2 B}p_{x_3}\left(p_{x_3} V_{x_2}-p_{x_2} V_{x_3}\right)\right) =0 \\
& V_{x_3}^{\prime \prime}+\left(-\frac{1}{z}-\phi^{\prime}+\frac{h^{\prime}}{h}+A_e^{\prime}+\frac{f^{\prime}}{f}+2B^{\prime}\right) V_{x_3}^{\prime} -\frac{ 2 g_5^{2} e^{2 A_e}}{f h z^2} M_{H}^{a a}(z) V_{x_3} \\ & +\frac{1}{f^2}\left(\omega \left(\omega V_{x_3}+p_{x_3} V_{t}\right)-f e^{-B} p_{x_1}\left(p_{x_1} V_{x_3}-p_{x_3} V_{x_1}\right)-f e^{- B}p_{x_2}\left(p_{x_2} V_{x_3}-p_{x_3} V_{x_2}\right)\right)  =0 .
\end{aligned}
\label{eomA}
\end{equation}

Since the components of the gauge field do not satisfy the Lorentz invariant, one can rewrite the equation of motion in terms of electric fields associated with the gauge potential, which is a Lorentz invariant quantity, 
\begin{equation}
E_i(p, z) \equiv \omega V_i(p, z)+p_i V_t(p, z),
\label{efield}
\end{equation}
By employing the following condition derived from the equation of motion,
\begin{equation}
\omega \partial_z V_t(p,z)+p^i \partial_z V_i(p, z)=0,
\end{equation}
one can establish a set of relations between the derivatives of the gauge field components with respect to the holographic axis ($z$) and the corresponding components of the electric field derivative with respect to $Z$.
\begin{equation}
\begin{aligned}
& \partial_z V_i=\frac{1}{\omega}\left(\delta_i^j-\frac{p_i p^j}{p^2}\right) \partial_z E_j \\
& \partial_z V_t=-\frac{p^i}{\omega^2}\left(\delta_i^j-\frac{p_i p^j}{p^2}\right) \partial_z E_j  .
\end{aligned}
\label{parA}
\end{equation}

Hence, by substituting Eq. \eqref{efield} and Eq. \eqref{parA} into Eq. \eqref{eomA}, we can write the equation of motions for the electric field $E_i$,
\begin{equation}
\begin{aligned}
& E_{x_1}^{\prime \prime}+\left(-\frac{1}{z}-\phi^{\prime}+\frac{h^{\prime}}{h}+A_e^{\prime}+\frac{f^{\prime}}{f}-B^{\prime}\right) E_{x_1}^{\prime} -\frac{ 2 g_5^{2} e^{2 A_e}}{f h z^2}M_{H}^{a a} E_{x_1} \\ & -(\frac{f^{\prime}}{f}-B^{\prime})  \left [ G_{x_1} p_{x_1}^{2} E_{x_1}^{\prime} + G_{x_2} p_{x_1} p_{x_2} E_{x_2}^{\prime} +G_{x_3} p_{x_1} p_{x_3} E_{x_3}^{\prime}  \right] + K E_{x_1} =0 \\
& E_{x_2}^{\prime \prime}+\left(-\frac{1}{z}-\phi^{\prime}+\frac{h^{\prime}}{h}+A_e^{\prime}+\frac{f^{\prime}}{f}-B^{\prime}\right) E_{x_2}^{\prime} -\frac{ 2 g_5^{2} e^{2 A_e}}{f h z^2}M_{H}^{a a} E_{x_2} \\ & -(\frac{f^{\prime}}{f}-B^{\prime})  \left [ G_{x_2} p_{x_2}^{2} E_{x_2}^{\prime} + G_{x_1} p_{x_2} p_{x_1} E_{x_1}^{\prime} +G_{x_3} p_{x_2} p_{x_3} E_{x_3}^{\prime}  \right] + K E_{x_2} =0 \\
& E_{x_3}^{\prime \prime}+\left(-\frac{1}{z}-\phi^{\prime}+\frac{h^{\prime}}{h}+A_e^{\prime}+\frac{f^{\prime}}{f}+2B^{\prime}\right) E_{x_1}^{\prime} -\frac{ 2 g_5^{2} e^{2 A_e}}{f h z^2}M_{H}^{a a} E_{x_3} \\ & -(\frac{f^{\prime}}{f}+2B^{\prime})  \left [ G_{x_3} p_{x_3}^{2} E_{x_3}^{\prime} + G_{x_1} p_{x_3} p_{x_1} E_{x_1}^{\prime} +G_{x_2} p_{x_3} p_{x_2} E_{x_3}^{\prime}  \right] + K E_{x_3} =0 ,
\end{aligned}
\label{eomE}
\end{equation}
where $G_{x_1}$, $G_{x_2}$, $G_{x_3}$, and $K$ are defined as the following,
\begin{equation}
\begin{aligned}
 G_{x_1(x_2)}&= \frac{g^{x_1 x_1 (x_2 x_2)}}{p^2}= \frac{f e^{-B}}{- \omega^2 + f e^{-B} (p_{x_1}^{2} + p_{x_2}^{2}) + f e^{2B} p_{x_3}^{2} },  \\
 G_{x_3}&= \frac{g^{x_3 x_3}}{p^2}= \frac{f e^{2B}}{- \omega^2 + f e^{-B} (p_{x_1}^{2} + p_{x_2}^{2}) + f e^{2B} p_{x_3}^{2} },  \\ 
 K &= \frac{1}{f^2} \left( \omega^2 - f e^{-B} (p_{x_1}^{2} + p_{x_2}^{2}) - f e^{2B} p_{x_3}^{2}  \right). \\
\end{aligned}
\label{GK}
\end{equation}

To calculate the retarded correlator, it is necessary to apply the correct boundary conditions at the horizon. To achieve this, we adopt the incoming boundary conditions at the horizon, located at $z=z_h$ \cite{Son:2002sd}. This allows us to derive the solutions near the horizon, which are expressed as follows: 
\begin{equation}
\begin{aligned}
& E_{x_1} (z)= (z_h -z)^{-i \frac{\omega}{4 \pi T}} \left( a_{x_1} + b_{x_1} (z_h-z) + \mathcal{O}[(z_h-z)^2]  \right), \\
& E_{x_2} (z)= (z_h -z)^{-i \frac{\omega}{4 \pi T}} \left( a_{x_2} + b_{x_2} (z_h-z) + \mathcal{O}[(z_h-z)^2]  \right), \\
& E_{x_3} (z)= (z_h -z)^{-i \frac{\omega}{4 \pi T}} \left( a_{x_3} + b_{x_3} (z_h-z) + \mathcal{O}[(z_h-z)^2]  \right), \\
\end{aligned}
\label{Ezh}
\end{equation}
where $a_{x_i}$ and $b_{x_i}$ are the integration constants. To find the two-point correlation function, one needs to take the functional derivative of the on-shell action with respect to the source of the gauge field. By substituting the equation of motion Eq. \eqref{emVm} into the action Eq. \eqref{actionnew}, we obtain the on-shell action  for the vector field
\begin{equation}
S_M=-\left. \int \frac{d^4 p}{(2 \pi)^4} \sqrt{-g} e^{-\phi(z)} \frac{h(\phi)}{2 g_{5}^{2} } g^{z z} g^{\mu \nu} V_\mu(-p, z) \partial_z V_\nu(p, z)\right|_{\epsilon} ^{z_h} .
\end{equation}

By expressing the action in terms of the components of the electric field, the following result can be derived:
\begin{equation}
S_M=\frac{-1}{2 g_{5}^{2} }\left. \int \frac{d^4 p}{(2 \pi)^4}  \frac{e^{-\phi(z)} f(z) h(\phi)}{z} \left[\frac{1}{\omega^2} (g^{i j} - \frac{p^i p^j}{p^2}) E_i (-p, z) \partial_z E_j (p, z)  \right] \right|_{\epsilon}^{z_h} .
\end{equation}

To derive the correlation functions, it is advantageous to decompose the electric field into a product of two components: the boundary value of the field, $E_i^{0} (p)$, and the bulk-to-boundary propagator, $\mathcal{E}_i(z)$, which is a function exclusively of the holographic coordinate ($z$),
\begin{equation}
E_i(z, k)=\mathcal{E}_i(z) E_i^{0} (p), \quad\left(i=x_1, x_2, x_3\right)
\end{equation}
where the function $\mathcal{E}_i(z)$ satisfy the condition $\lim _{z \rightarrow \epsilon} \mathcal{E}_i(z)=1$. Now, the on-shell action becomes 
\begin{equation}
S_M=\frac{-1}{2 g_{5}^{2} }\left. \int \frac{d^4 p}{(2 \pi)^4}  \frac{e^{-\phi(z)} f(z) h(\phi)}{z} \frac{1}{\omega^2} (g^{i j} - \frac{p^i p^j}{p^2}) \mathcal{E}_i(z) \partial_z \mathcal{E}_j(z) \left( \omega^{2} V_{i}^{0} V_{j}^{0} + \omega p_{j} V_{i}^{0} V_{t}^{0} + \omega p_{i} V_{t}^{0} V_{j}^{0} + p_{i} p_{j} V_{t}^{0} V_{t}^{0}\right) \right|_{\epsilon}^{z_h},
\end{equation}
where one can use a compact form which reduces it to the following 
\begin{equation}
S_M=\left. \int \frac{d^4 p}{(2 \pi)^4}  V_{\mu}^{0}(-p) \mathcal{F}^{\mu \nu} (z, p) V_{\nu}^{0}(p) \right|_{\epsilon}^{z_h},
\end{equation}
with the components of the function $\mathcal{F}^{\mu \nu}$ are given by
\begin{equation}
\begin{aligned}
& \mathcal{F}^{t t} (z, p)= \frac{-1}{2 g_{5}^{2} } \frac{e^{-\phi(z)} f(z) h(\phi)}{z} \frac{1}{\omega^2} (g^{i j} - \frac{p^i p^j}{p^2}) \mathcal{E}_i(z) \partial_z \mathcal{E}_j(z) \left(p_{i} p_{j}\right) , \\
& \mathcal{F}^{i j} (z, p)= \frac{-1}{2 g_{5}^{2} } \frac{e^{-\phi(z)} f(z) h(\phi)}{z} \frac{1}{\omega^2} (g^{i j} - \frac{p^i p^j}{p^2}) \mathcal{E}_i(z) \partial_z \mathcal{E}_j(z) \left(\omega^2\right) , \\
& \mathcal{F}^{i t} (z, p)= \frac{-1}{2 g_{5}^{2} } \frac{e^{-\phi(z)} f(z) h(\phi)}{z} \frac{1}{\omega^2} (g^{i j} - \frac{p^i p^j}{p^2}) \mathcal{E}_i(z) \partial_z \mathcal{E}_j(z) \left(\omega p_{j}\right),\\
& \mathcal{F}^{t j} (z, p)= \frac{-1}{2 g_{5}^{2} } \frac{e^{-\phi(z)} f(z) h(\phi)}{z} \frac{1}{\omega^2} (g^{i j} - \frac{p^i p^j}{p^2}) \mathcal{E}_i(z) \partial_z \mathcal{E}_j(z) \left(\omega p_{i}\right).
\end{aligned}
\label{Fmunu}
\end{equation}

The two-point correlation function using the Son-Starinets prescription \cite{Son:2002sd} can be defined as
\begin{equation}
D^{\mu \nu}(p)= 2 \lim _{z \rightarrow \epsilon} \mathcal{F}^{\mu \nu}(z, p) .
\end{equation}

\subsection{Spin density matrix}

The spin state of a vector meson is described by a $3 \times 3$ Hermitian spin-density matrix.  The elements of the spin-density matrix can be studied by measuring the  angular distributions of the decay products of vector mesons with reference to a quantization axis. One of the decay channels to investigate the spin alignment of the vector meson is through the decay to the dilepton. For the two-body dilepton decay of a vector meson, the angular distribution is given by \cite{Faccioli:2010kd}
\begin{equation}
W(\theta) \propto \frac{1}{3+\lambda_\theta}\left(1+\lambda_\theta \cos ^2 \theta\right),
\end{equation}
where $\theta$ is the polar emission angle of the positively charged decay lepton regarding to a chosen quantization direction. For dilepton decays, the $\lambda_\theta$ parameter is related to $\rho_{00}$ through the following relation,
\begin{equation}
\lambda_\theta=\frac{1-3 \rho_{00}}{1+\rho_{00}}
\end{equation}
where the deviation of $\lambda_\theta$ from zero, which corresponds to the $\rho_{00} \neq 1/3$, is a signal of the alignment of a vector meson. The S-matrix element for a vector meson decays to a muon pair in the final state is
\begin{equation}
\mathcal{M}_{f i}=\int d^4 x d^4 y\langle f, l \bar{l}| J_\mu(y) G_R^{\mu \nu}(x-y) J_\nu^l(x)|i\rangle ,
\end{equation}
where $J_\mu(y)$ and $J_\nu^l$ are the hadronic and leptonic current respectively. $G_R^{\mu \nu}$ is the retarded propagator of the vector meson at the vacuum and is defined by
\begin{equation}
G_R^{\mu \nu}(p)=-\frac{\eta^{\mu \nu}+p^\mu p^\nu / p^2}{p^2+m_{V}^2+i m_{V} \Gamma},
\end{equation}
with $m_{V}$ the vacuum mass and $\Gamma$ is the width of the vector meson.  By summing over all possible final states and taking the average over the initial state, one can derive the total production rate of the dimuons \cite{Sheng:2024kgg}.
\begin{equation}
n(x, p)=  -\frac{2 g_{M \mu^+ \mu^-}^2}{3(2 \pi)^5}\left(1-\frac{2 m_{\mu}^2}{p^2}\right) \sqrt{1+\frac{4 m_{\mu}^2}{p^2}} p^2 n_B(x, \omega)  \times\left(\eta_{\mu \nu}+\frac{p_\mu p_\nu}{p^2}\right) G_A^{\mu \alpha}(p) \varrho_{\alpha \beta}(x, p) G_R^{\beta \nu}(p),
\label{productionrate}
\end{equation}
where $n_B(x, \omega)=$ $1 /\left[e^{\omega / T(x)}-1\right]$ is the Bose-Einstein distribution, the muon mass $m_{\mu} = 0.105$ GeV, $g_{M \mu^+ \mu^-}$ being the coupling strength, $G_R(A)^{\mu \nu}$ retarded (advanced) propagator at the vacuum, and
\begin{equation}
\varrho_{\alpha \beta}(x, p) \equiv - \operatorname{Im} D_{\alpha \beta}(x, p)
\end{equation}
is the spectral function in the medium, which is defined by the imaginary part of the retarded current-current correlator. The spectral function can be expressed through a decomposition using a complete set of polarization vectors to obtain the separate contributions from different spin states.
\begin{equation}
\varrho^{\mu \nu}(x, p)=\sum_{\lambda, \lambda^{\prime}=0, \pm 1} v^\mu(\lambda, p) v^{* \nu}\left(\lambda^{\prime}, p\right) \tilde{\varrho}_{\lambda \lambda^{\prime}}(x, p),
\label{spectral}
\end{equation}
where $v^\mu(\lambda, p)$ is the covariant form of spin polarization vectors which is given by
\begin{equation}
v^\mu(\lambda, p)=\left(\frac{\mathbf{p} \cdot \mathbf{\epsilon}_\lambda}{M}, \boldsymbol{\epsilon}_\lambda+\frac{\mathbf{p} \cdot \mathbf{\epsilon}_\lambda}{M(\omega+M)} \mathbf{p}\right).
\end{equation}
with $M \equiv \sqrt{\omega^2-\mathbf{p}^2}$ the invariant mass for the vector meson. The polarization vector satisfies both the orthonormality and completeness conditions such that
\begin{equation}
\begin{aligned}
& \eta_{\mu \nu} v^\mu(\lambda, p) v^{* \nu}\left(\lambda^{\prime}, p\right)=\delta_{\lambda \lambda^{\prime}} \\
& \sum_\lambda v^\mu(\lambda, p) v^{* \nu}(\lambda, p)=\left(\eta^{\mu \nu}+p^\mu p^\nu / p^2\right).
\end{aligned}
\end{equation}

The three-component vectors $\mathbf{\epsilon}_\lambda$ represent the three possible spin orientations in the meson's rest frame, with $\mathbf{\epsilon}_0$ corresponds to the direction of spin quantization, and $\mathbf{\epsilon}_{ \pm 1}$ being perpendicular to $\mathbf{\epsilon}_0$. By using the spectral function in the spin space, which can be
extracted from Eq. \eqref{spectral} \footnote{ for the details of the explicit form of the spectral function in the spin space see Appendix \ref{sfss}},
\begin{equation}
\tilde{\varrho}_{\lambda \lambda^{\prime}}(p)=\eta_{\mu \alpha} \eta_{\nu \beta} v^{* \alpha}(\lambda, p) v^\beta\left(\lambda^{\prime}, p\right) \varrho^{\mu \nu}(p)
\label{spinspectral}
\end{equation}

Then it is possible to write Eq. \eqref{productionrate} in terms of different spin states $\lambda=0, \pm 1$,
\begin{equation}
n_\lambda(x, p)=  -\frac{2 g_{M\mu^+ \mu^-}^2}{3(2 \pi)^5}\left(1-\frac{2 m_{\mu}^2}{p^2}\right) \times \sqrt{1+\frac{4 m_{\mu}^2}{p^2}}  \frac{p^2 n_B(x, \omega) \tilde{\varrho}_{\lambda \lambda}(x, p)}{\left(p^2+m_V^2\right)^2+m_V^2 \Gamma^2}  .
\end{equation}

The total dilepton number is the summation of all spin states, $n=\sum_{\lambda=0, \pm 1} n_\lambda$. The spin alignment for the vector meson is defined as the probability at a particular spin state, and for the produced dimuon is defined as
\begin{equation}
\rho_{\lambda \lambda^{\prime}}(p)=    -\frac{2 g_{M\mu^+ \mu^-}^2}{3(2 \pi)^5 N}\left(1-\frac{2 m_{\mu}^2}{p^2}\right) \times \sqrt{1+\frac{4 m_{\mu}^2}{p^2}}  \frac{p^2 n_B(x, \omega) \tilde{\varrho}_{\lambda \lambda^{\prime}}(x, p)}{\left(p^2+m_V^2\right)^2+m_V^2 \Gamma^2},
\label{spindens}
\end{equation}
with $N$ being the normalization factor that ensures the matrix is properly normalized ($N=\sum_{\lambda=0, \pm 1} \rho_{\lambda \lambda}$), which guarantees that the sum of the diagonal part of the spin state is unity. Note that for the restricted invariant mass near to the resonance mass $-p^2=m_V^2$, the spin alignment can be approximate to the ratio $\tilde{\varrho}_{\lambda \lambda^{\prime}} / \sum_\lambda \tilde{\varrho}_{\lambda \lambda}$.

\section{Results}
\label{sectionIII}

By utilizing the introduced geometry, we are now prepared to investigate the spin alignment of the $\rho$, $\phi$, and $J/\Psi$ vector mesons within the four-flavor soft-wall holographic QCD framework. Initially, we will examine the confinement/deconfinement transition temperature in the presence of rotation and chemical potential. Following that, we will analyze the spectral functions of the vector mesons and determine their melting points at finite temperature and angular velocity. Ultimately, we aim to address the primary focus of our work: the impact of rotation on the spin alignment of the $\rho$, $\phi$, and $J/\Psi$ mesons in a hot medium. 

In this study, we perform the analysis where the spin quantization direction is along the event plane (the $x_2$-direction), which, in heavy ion collision, is known as the event plane frame. It is important to note that the model consists of five parameters: $\mu_G$, $\mu_{G_2}$, $\mu_{\Omega}$, $h_s$, and $h_c$. The parameters $\mu_G$, $h_s$, and $h_c$ are determined from the study of the meson spectra at zero temperature. Specifically, $\mu_{G}$ is fitted to be $0.43$ GeV based on the higher excited states of the $\rho$ meson masses, while $h_s$ and $h_c$ are set to $0.10$ GeV and $0.45$ GeV, respectively, by fitting the masses of the $\phi$ and $J/\Psi$ mesons. The value of $\mu_{G_2}$ is selected to be large ($\mu_{G_2}=3$) to ensure it does not influence the IR physics \cite{Li:2014hja}. Lastly, $\mu_{\Omega}$ is assigned a value of $10$ GeV$^{-1}$ to achieve a qualitative correspondence of the critical temperature as a function of angular velocity, similar to lattice QCD results \cite{Braguta:2021jgn}.

\subsection{Temperature vs rotation}

To incorporate the effects of rotating QCD matter, we introduced anisotropic backgrounds that can be obtained by solving the EMD action. The effects of rotation are introduced into these anisotropic backgrounds through an Abelian gauge field and a scalar (dilaton) field, which are related to the conserved $U(1)$ current and the gluonic operator at the boundary of the geometry, respectively. Additionally, a non-zero baryon chemical potential is introduced via the $U(1)$ gauge field, as explained in the previous section. Since there are no contributions from quark flavors in our background, it remains purely gluonic. For such a background system at zero chemical potential, one would expect the confinement/deconfinement transition to be of first order.

\begin{figure}
  \centering
  \includegraphics[width=0.47\linewidth]{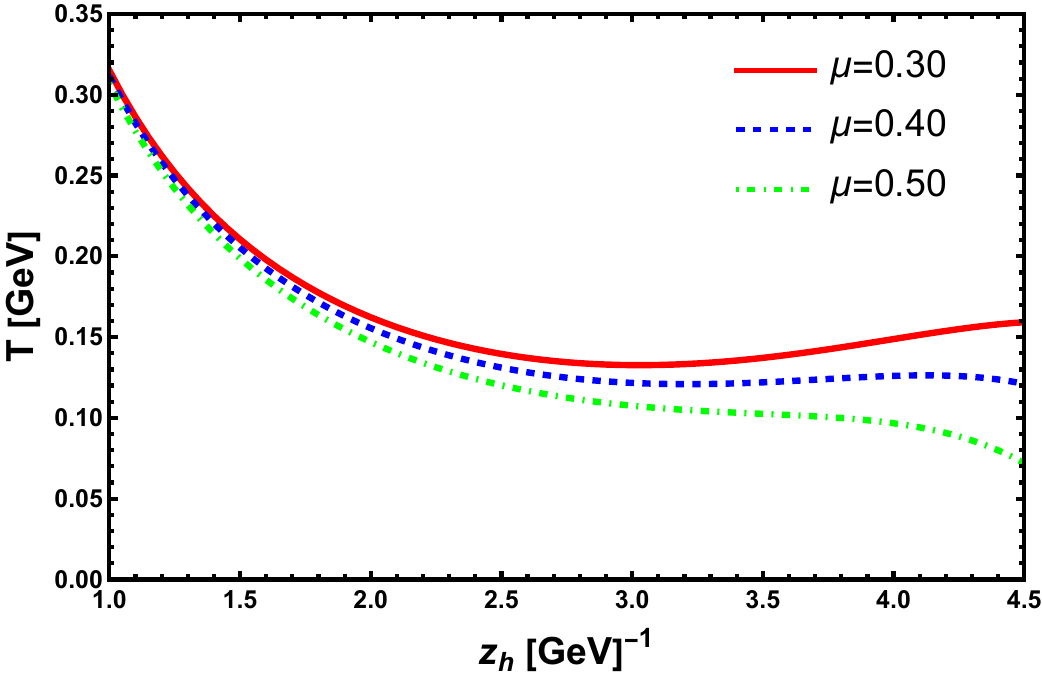}
  \includegraphics[width=0.49\linewidth]{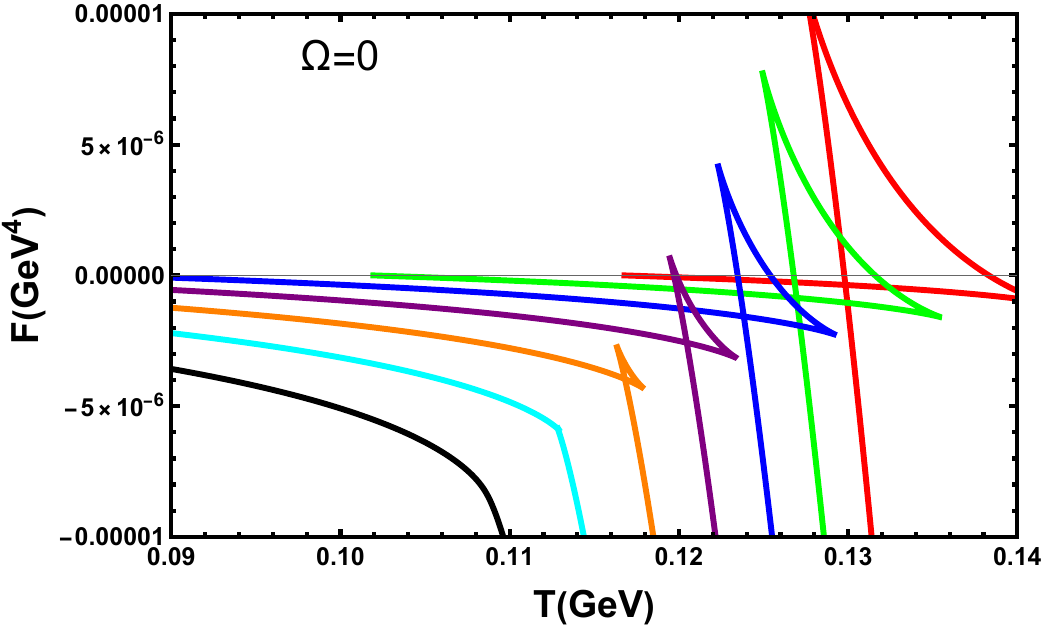}
\caption{ Left: The temperature as a function of $z_h$ for finite values of the chemical potential $\mu=0.30$ GeV (solid red line), $\mu=0.40$ GeV (dashed blue line), and $\mu=0.50$ GeV (dot-dashed green line). Right: The free energy density as a function of temperature at different values of the chemical potential, which starts from $\mu=0.35$ GeV (Solid red line) to $\mu=0.47$ GeV (Solid black line) with the step size of $0.02$ GeV.}
\label{tempmu}
\end{figure}

The temperature of the hot medium, as a function of the black hole horizon, is described by Eq. \eqref{temp}. The behavior of the temperature at finite chemical potential is illustrated in Fig. \ref{tempmu}. As shown in the left panel of Fig. \ref{tempmu}, the behavior of the temperature depends on the value of the chemical potential. In particular, when $\mu=0.30$ GeV, there is a global minimum, after which the temperature increases again. This minimum point separates two distinct phases: the large black hole phase (stable) and the small black hole phase (unstable). As the chemical potential increases to $\mu=0.45$ GeV, the unstable phase disappears, resulting in a plateau region in the ($T-z_h$) diagram for a finite horizon value. Beyond this chemical potential, the temperature consistently declines. Notably, the disappearance of the unstable phase signifies the presence of a critical end point (CEP) in the $T-\mu$ diagram. 

We need to calculate the free energy density to determine the precise location of the CEP. The free energy as a function of temperature at different chemical potential values is presented in the right panel of Fig. \ref{tempmu}. The swallowtail behavior of the free energy density indicates a first-order phase transition, while the disappearance of this behavior signals a change to crossover. The point where the order of the phase transition changes from first order to crossover marks the CEP, which is located at ($T_{CEP}, \mu_{CEP}) = (0.1128, 0.45)$ GeV.

\begin{figure}
  \centering
  \includegraphics[width=0.49\linewidth]{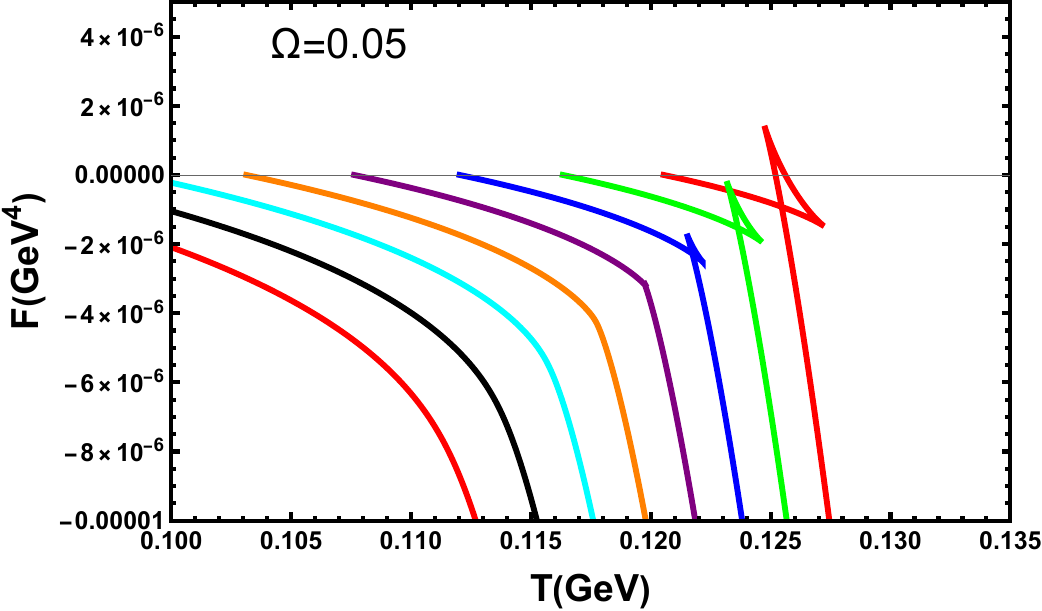} 
  \includegraphics[width=0.49\linewidth]{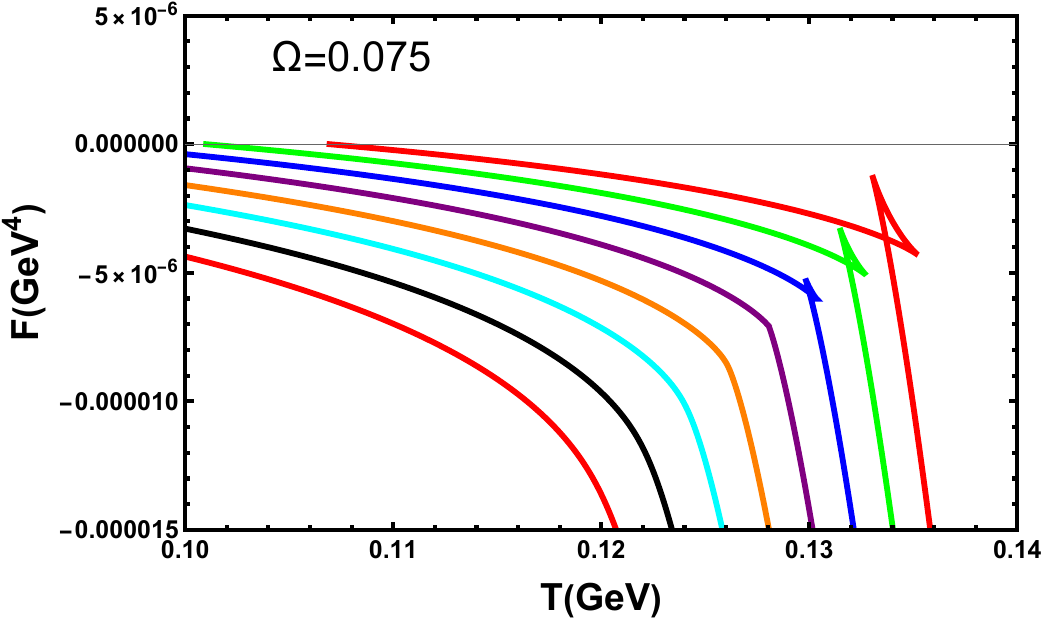}   
\caption{The free energy density as a function of temperature at different values of the chemical potential, which starts from $\mu=0.43$ GeV (Solid red line) to $\mu=0.50$ GeV (Solid black line) with the step size of $0.01$ GeV (left panel), and from $\mu=0.45$ GeV (Solid red line) to $\mu=0.52$ GeV (Solid black line) with the step size of $0.01$ GeV (right panel).   }
\label{freeT5}
\end{figure}

Then a question arises what is the effect of the rotation on the location of the CEP? For that reason, we show the free energy as a function of temperature at finite angular velocity and different chemical potential values in Fig. \ref{freeT5}. As shown in Fig. \ref{freeT5}, increasing the angular velocity shifts the CEP to a higher temperature and chemical potential, such that by moving from $\Omega=0$ to $\Omega=0.05$ GeV, the CEP changes from ($T_{CEP}, \mu_{CEP}) = (0.1128, 0.45)$ GeV to ($T_{CEP}, \mu_{CEP}) = (0.1195, 0.46)$ GeV, and at $\Omega=0.075$ GeV, it moves to ($T_{CEP}, \mu_{CEP}) = (0.128, 0.48)$. Then it is obvious that the effect of the rotation is opposite to the chemical potential.

\begin{figure}
  \centering
  \includegraphics[width=0.47\linewidth]{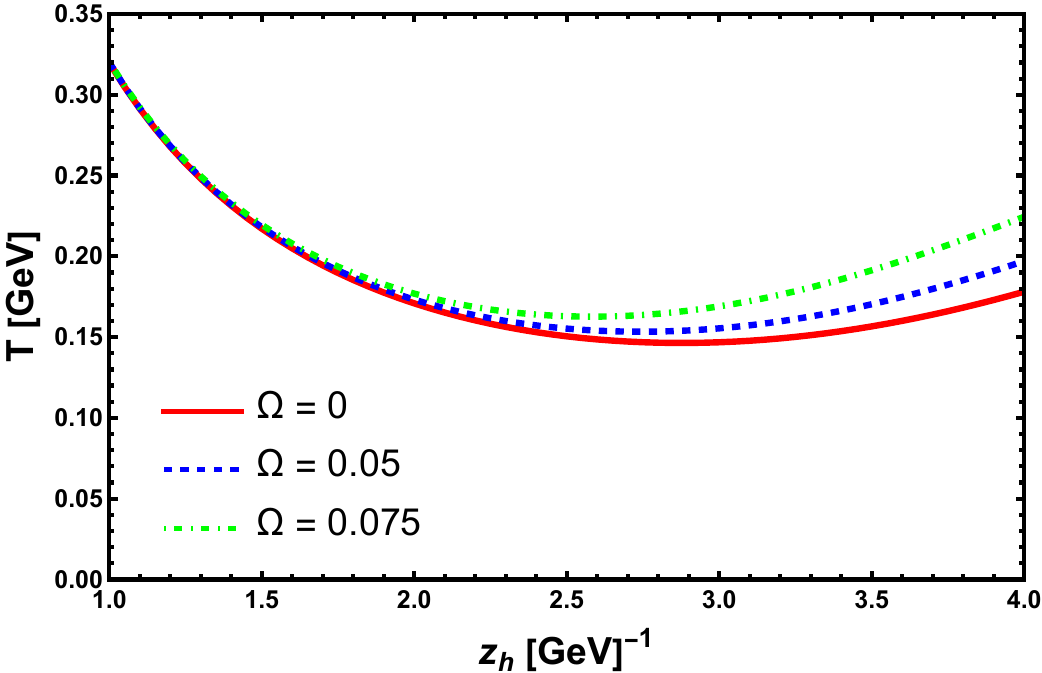}
\caption{The temperature as a function of black hole horizon at different values of the angular velocity $\Omega=0.0$ GeV (solid red line), $\Omega=0.05$ GeV (dashed blue line), and $\Omega=0.075$ GeV (dot-dashed green line).}
\label{tempomega}
\end{figure}

The primary focus of this work is to examine the effect of a rotating background on observables. From this point onward, we will consider the case of zero chemical potential. It is crucial to note that this approach is valid only under the near-centre approximation \cite{Chen:2024jet}. Fig. \ref{tempomega} illustrates how temperature behaves as a function of the black hole horizon at finite angular velocity ($\Omega$). Unlike the chemical potential, the effect of rotation does not alter the order of the confinement/deconfinement transition, which remains first-order. Additionally, the global minimum of the free energy shifts to a higher temperature as the angular velocity increases. This increasing behavior of the transition temperature in a rotating medium is consistent with recent LQCD findings \cite{Braguta:2021jgn}.

\begin{figure}
  \centering
 \includegraphics[width=0.505\linewidth]{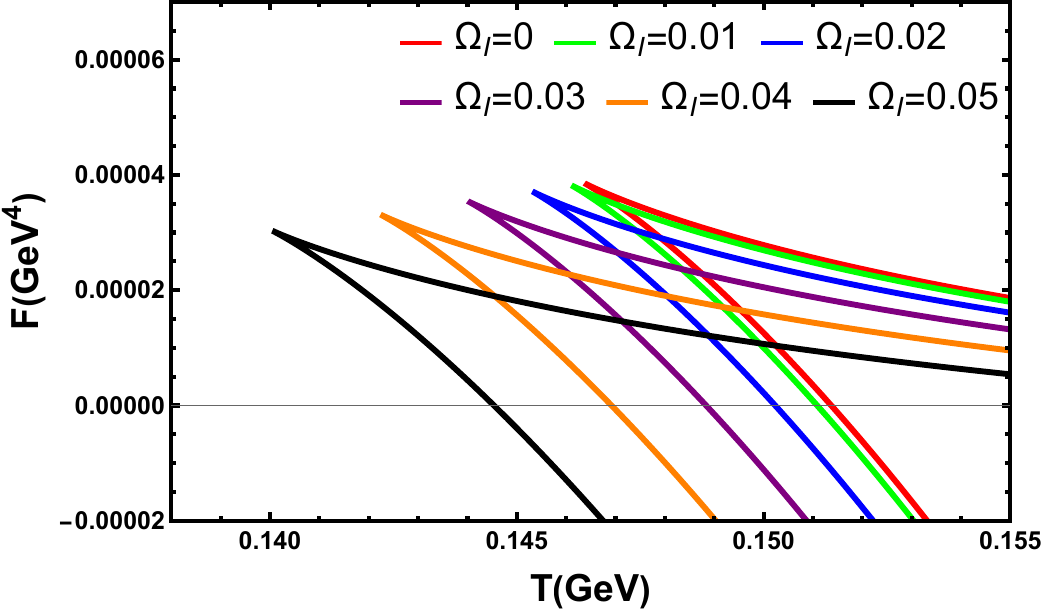} 
  \includegraphics[width=0.475\linewidth]{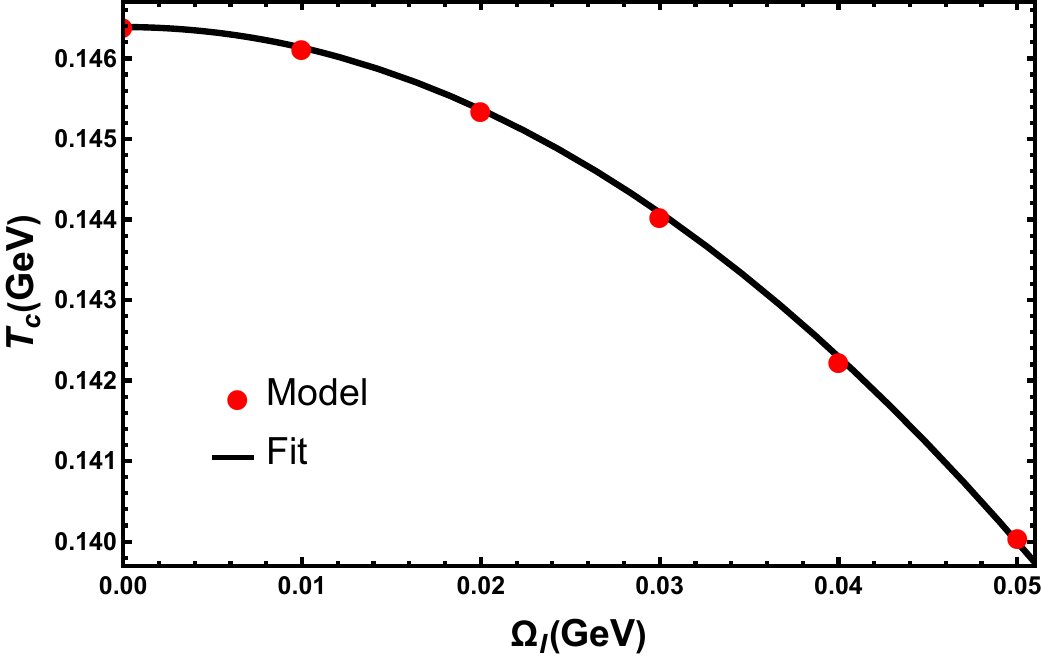}   

\caption{The left panel shows the free energy density as a function of temperature at different values of the imaginary angular velocity. The right panel is the $T-\Omega_I$ diagram, where the red dot is the model result, and the black line is the simple fitting function.}
\label{freeT}
\end{figure}

As previously mentioned, the source of rotation in the medium arises from the dilaton field and the gauge field. 
To qualitatively compare our results with LQCD \cite{Braguta:2021jgn,Braguta:2022str}, we present the results of the free energy density and critical temperature using the imaginary angular velocity in Fig. \ref{freeT} \footnote{LQCD faces the sign problem in the rotating medium. To overcome this issue the imaginary angular velocity has been used in Refs. \cite{Braguta:2021jgn,Braguta:2022str}.}. As shown in Fig. \ref{freeT}, the transition temperature is determined from the swallowtail of the free energy density. The critical temperature of the confinement/deconfinement phase transition ($T_c$) shifts to the lower value at finite imaginary angular velocity. The behavior of $T_c$ with $\Omega_I$ can be fitted with a simple function such that 
\begin{equation}
    \frac{T_{c}(\Omega_I)}{T_{c}(0)}=1 - C ~ \Omega_{I}^{2}.
\end{equation}

This behaviour aligned with the LQCD and suggests using a small value of the angular velocities. By naively extrapolating the imaginary rotation to the real angular velocity, we conclude that the critical temperature of the confinement/deconfinement transition increases as the angular velocity rises.

\begin{figure}
  \centering
  \includegraphics[width=0.49\linewidth]{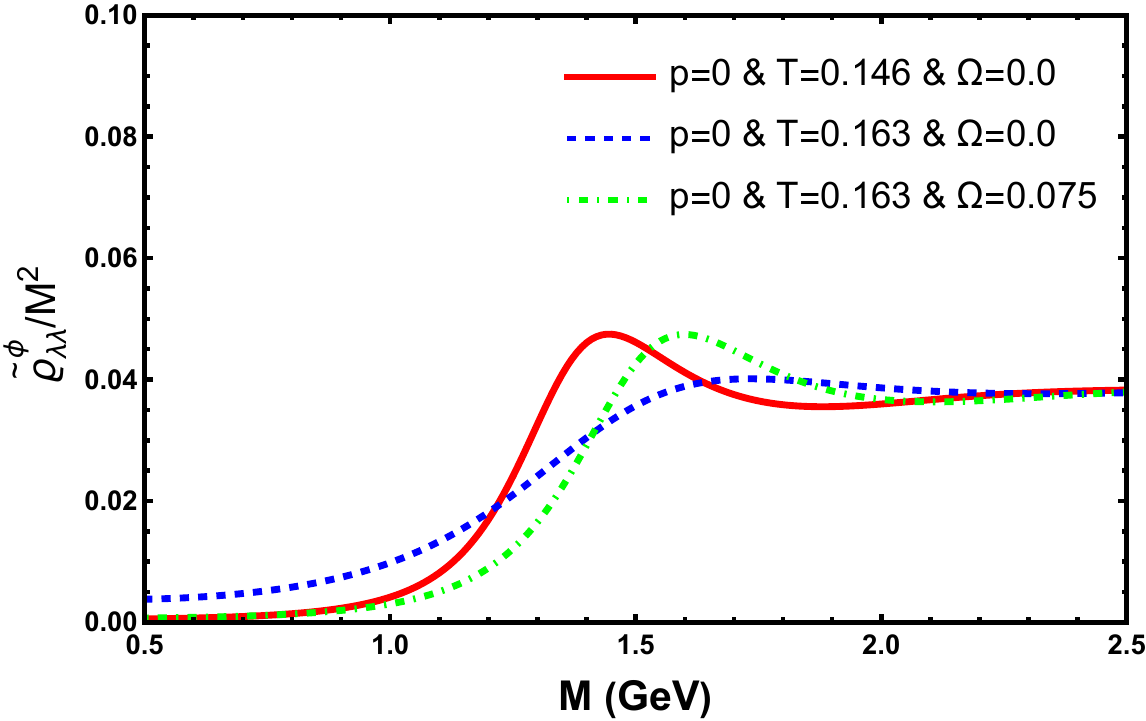} 
  \includegraphics[width=0.49\linewidth]{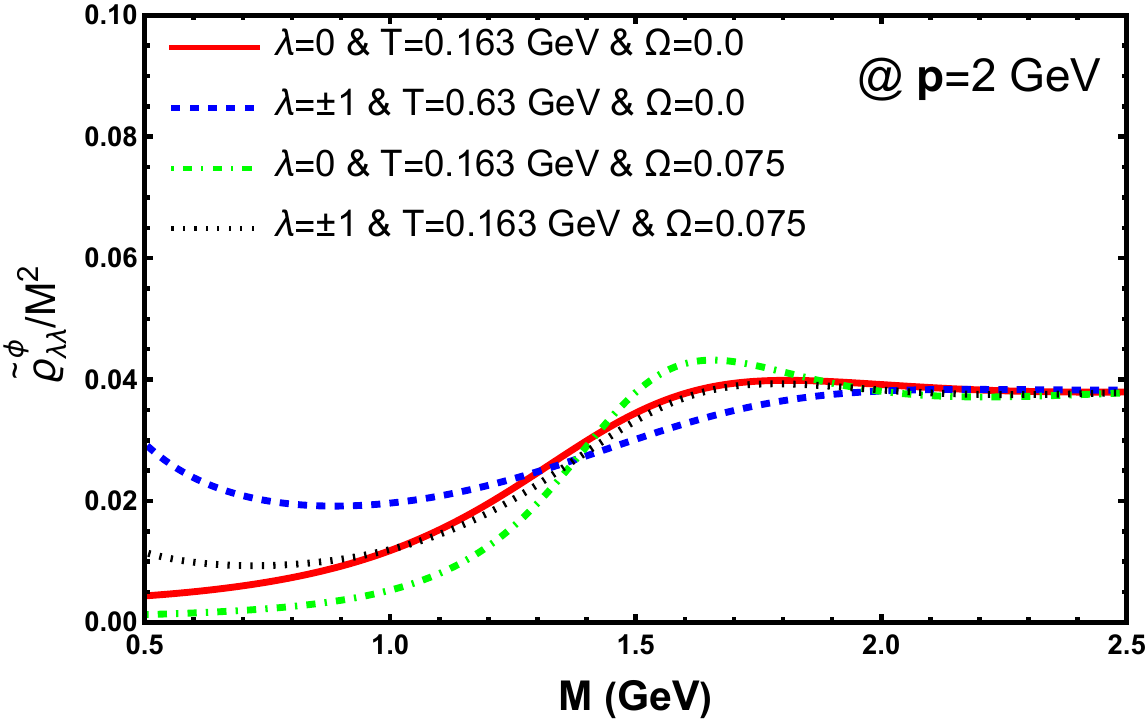}   
\caption{ Spectral functions $\tilde{\varrho}_{\lambda \lambda}$ of the $\phi$ meson as a function of the invariant mass $M$. In the left panel, the longitudinal ($\lambda=0$) component of the spectral function is given at $\textbf{p}=0$, $\Omega=0$, and $T=0.146$ GeV (solid red line) and $T=0.163$ GeV (dashed blue line, and at $\Omega=0.0.075$ GeV, and $T=0.163$ GeV (dot-dashed green line). Right panel: The longitudinal ($\lambda=0$) and transverse ($\lambda=\pm 1$) components of the spectral function are given at $\textbf{p}=2$ GeV, $\Omega=0$, and $T=0.168$ GeV, solid red and dashed blue line, respectively, and $\textbf{p}=2$ GeV, $\Omega=0.075$ GeV, and $T=0.168$ GeV, dot-dashed green and dotted black line, respectively.   }
\label{spectralphi}
\end{figure} 

\begin{figure}
  \centering
  \includegraphics[width=0.49\linewidth]{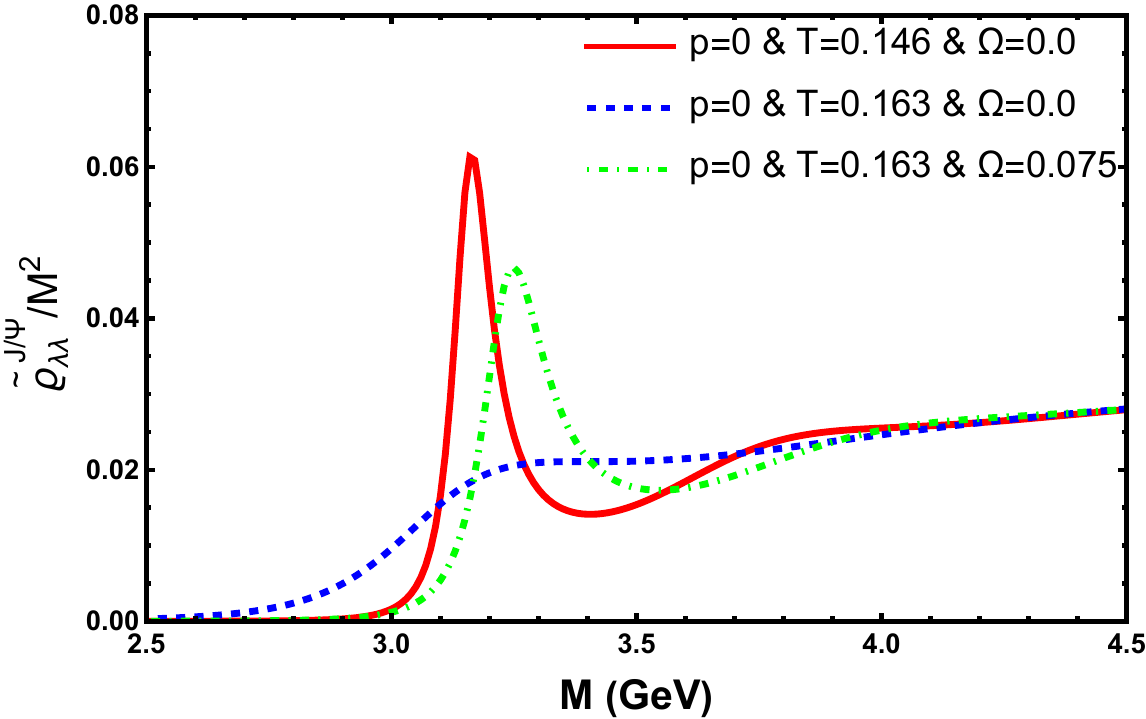} 
  \includegraphics[width=0.49\linewidth]{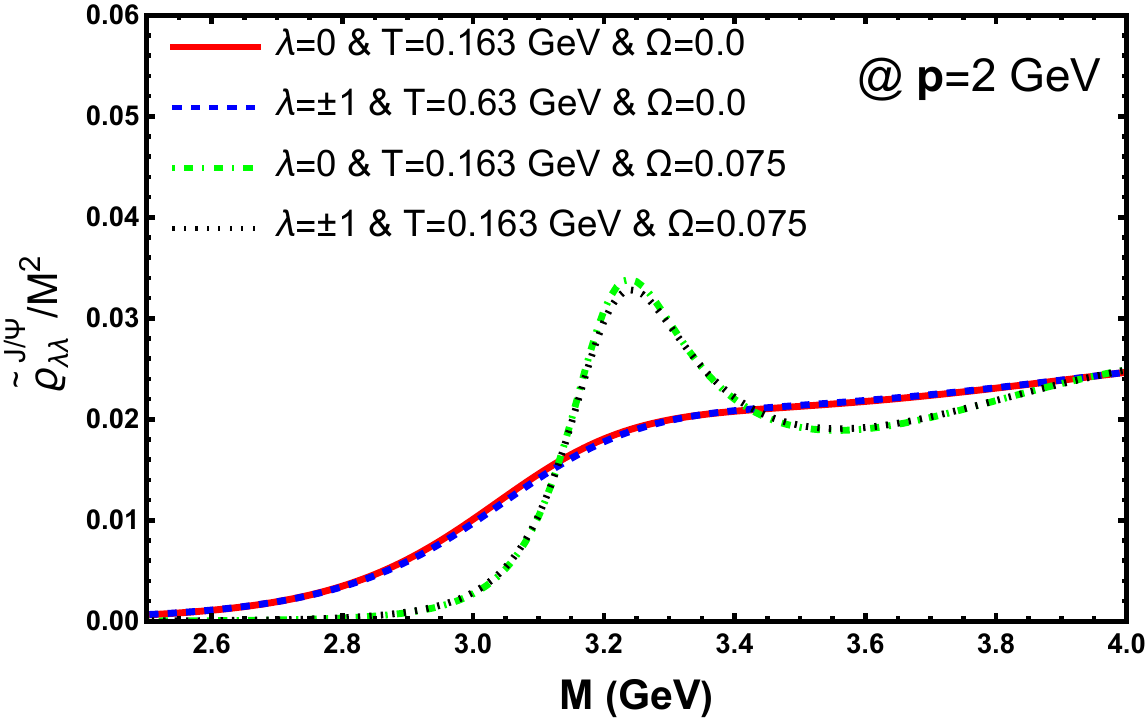}   
\caption{Spectral functions $\tilde{\varrho}_{\lambda \lambda}$ of the $J/\Psi$ meson as a function of the invariant mass $M$. The color online is similar to Fig. \ref{spectralphi}.  }
\label{spectraljpsi}
\end{figure} 

\begin{figure}
  \centering
  \includegraphics[width=0.49\linewidth]{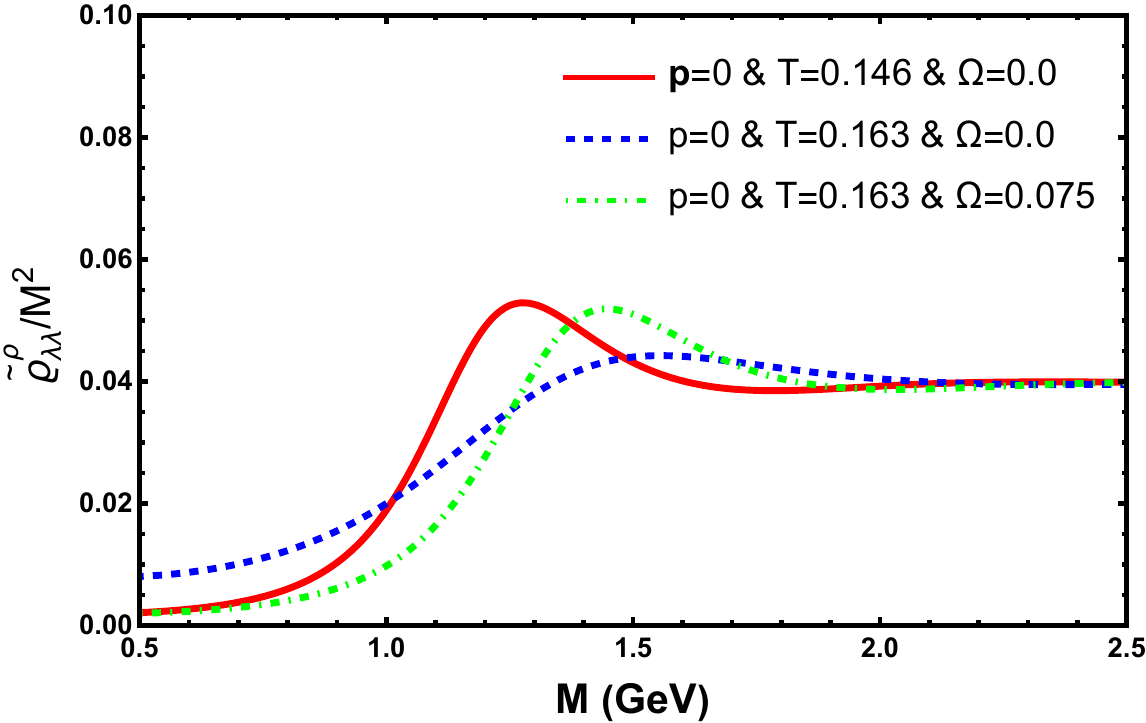} 
  \includegraphics[width=0.49\linewidth]{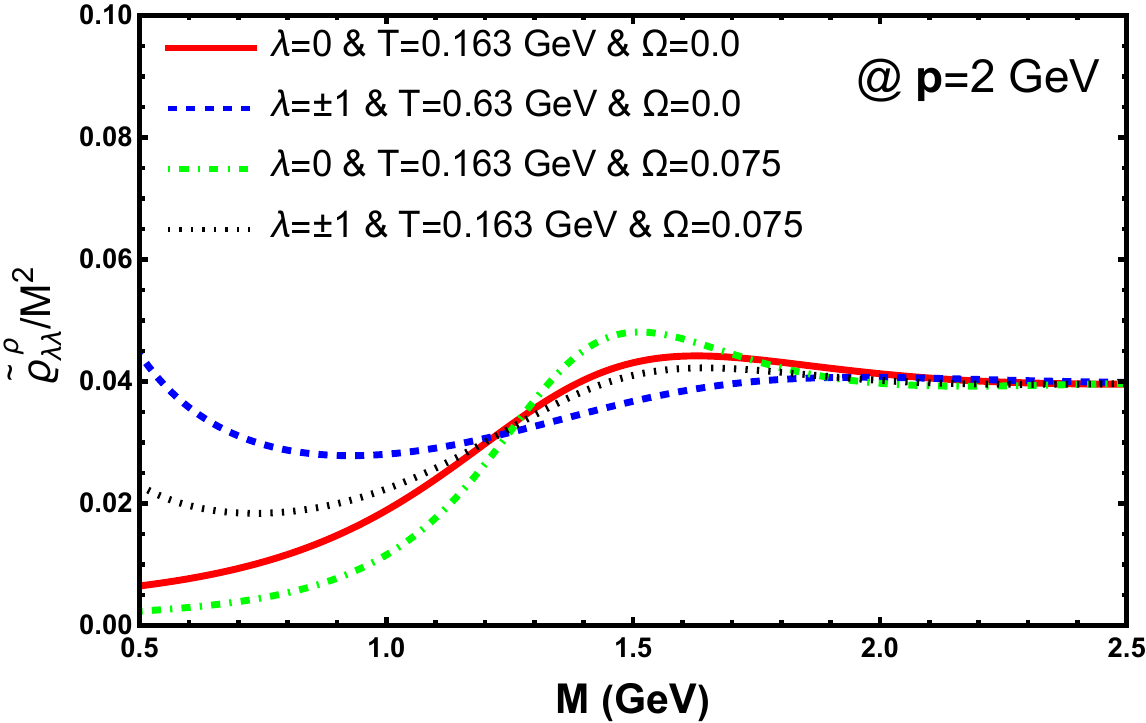}   
\caption{Spectral functions $\tilde{\varrho}_{\lambda \lambda}$ of the $\rho$ meson as a function of the invariant mass $M$. The color online is similar to Fig. \ref{spectralphi}.  }
\label{spectralrho}
\end{figure} 

\subsection{Melting of vector mesons }

To study the vector mesons ($\rho$, $\phi$, $J/\Psi$) in a rotating anisotropic background, it is necessary to introduce an $SU(4)_V$ gauge field as a probe in this context. We examine the spectral functions for any peaks to identify vector mesons in the medium. The positions of these peaks on the horizontal axis indicate the meson's mass, while the width of the peaks is associated with the thermal decay width of the vector meson. The point at which a peak in the spectral functions disappears signals the melting of the vector meson in the medium.

The spectral functions of the $\phi$ meson in the spin space, as a function of the invariant mass ($M$), are illustrated in Fig. \ref{spectralphi}. In the left panel, we present the results for the longitudinal spectral function ($\lambda =0$) at zero spatial momentum, along with varying temperature and angular velocity. It is important to note that, at zero momentum, the spectral functions for all spin states exhibit slight differences due to the effects of rotation; however, these differences do not affect the location of the peak. 

For a temperature of $T=0.146$ GeV and zero angular velocity, a broad peak appears in the spectral function, which can be interpreted as the presence of the $\phi$ meson. When the temperature is increased to $T=0.163$ GeV, this peak disappears, indicating that the $\phi$ meson has melted in the medium. Interestingly, the broad peak reemerges in the spectral function when the temperature is held at $T=0.163$ GeV while introducing an angular velocity of $\Omega =0.075$ GeV. This observation suggests that vector mesons in a rotating medium dissociate at a higher temperature compared to a medium without rotation.

In the right panel of Fig. \ref{spectralphi}, we display the longitudinal ($\lambda =0$) and transverse ($\lambda =\pm 1$) components of the spectral function at a spatial momentum of $\boldsymbol{p}=2$ GeV, with a temperature of $T=0.163$ GeV, and for both rotating and non-rotating cases. A significant distinction is observed between the spectral functions for the longitudinal and transverse components for both conditions. Additionally, compared to the zero momentum scenario, the peak values of the spectral functions decrease, reflecting that high-momentum resonances are more challenging to produce than low-momentum resonances.

Moreover, we present the spectral functions of the $J/\Psi$ meson in spin space as a function of invariant mass, as shown in Fig. \ref{spectraljpsi}. In contrast to the $\phi$ meson, a prominent peak appears in the spectral function of the $J/\Psi$ meson, which can be interpreted as a quasi-particle state. At zero spatial momentum, this peak disappears in the hot medium at a temperature of $T=0.163$ GeV. However, when introducing a rotational effect in the medium with $\Omega=0.075$ GeV, the $J/\Psi$ meson persists. Although the rotation slightly disrupts the symmetry between the longitudinal and transverse components of the spectral function, for a spatial momentum of $\textbf{p}=2$ GeV, the separation between these two components is not as pronounced as it is for the $\phi$ meson.

Finally, the spectral functions of the $\rho$ meson as a function of the invariant mass $M$ are shown in Fig. \ref{spectralrho}. The qualitative behaviour of the $\rho$ meson is similar to $\phi$, with the only difference being the position of the peak in a smaller invariant mass.

\subsection{Spin alignment of vector mesons}

The components of the spin density matrix for a spin one particle can be obtained from Eq.  \eqref{spindens}. Here, we are interested to investigate the spin alignment of the $\phi$, $J/\Psi$, and $\rho$ mesons. In heavy ion collisions, the momentum of the meson is parameterized by the transverse momentum $p_T$, azimuthal angle $\varphi$, and rapidity $y$, such that
\begin{equation}
    p^{\mu}=\left( \sqrt{p_{T}^{2} + M^{2}} \cosh(y) , p_T \sin(\varphi), p_T \cos(\varphi) , \sqrt{p_{T}^{2} + M^{2}} \sinh(y) \right).
\end{equation}

Recall that the $x_2$-axis in the meson's rest frame is chosen as the direction of the spin quantization. We begin by examining the spin alignment signal of the $\phi$ meson. As demonstrated by the experimental collaboration \cite{STAR:2022fan}, it is only possible to measure the spin alignment parameter \( \rho_{00} \). It is significant to mention that when we consider a thermal non-rotating background, the minimum temperature we can achieve is $T=0.146$ GeV. In contrast, with rotation at $\Omega=0.075$ GeV, the minimum temperature rises to $T=0.163$ GeV.

\begin{figure}
  \centering
  \includegraphics[width=0.55\linewidth]{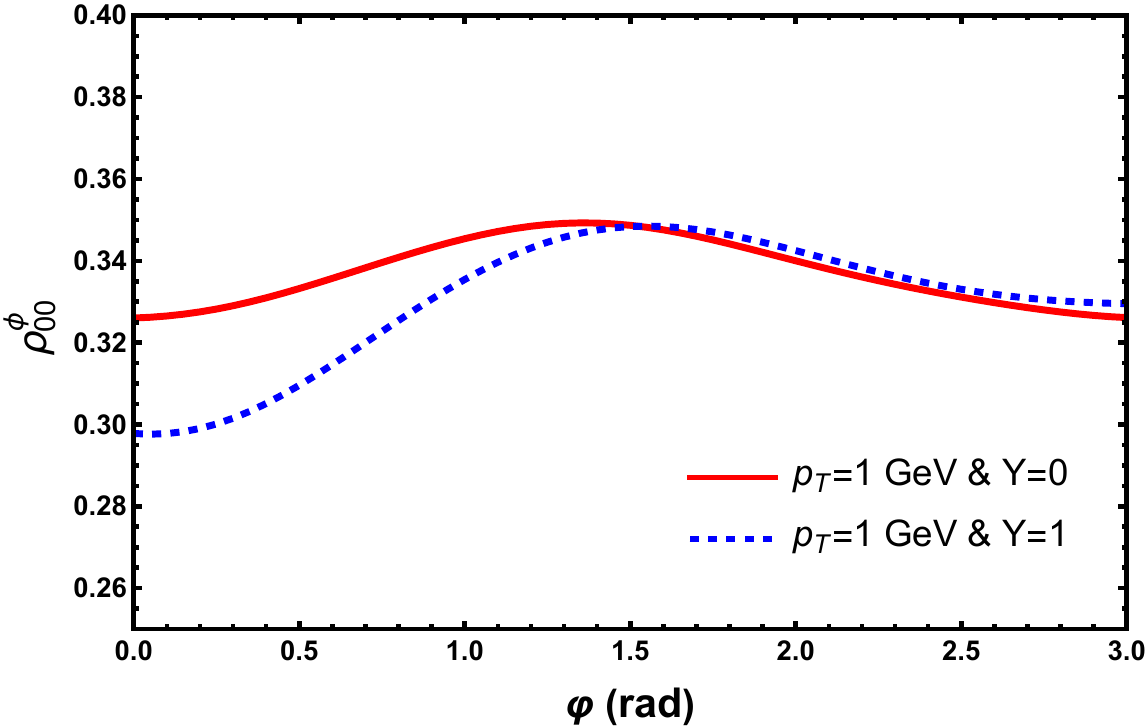} 
\caption{ The $\rho_{00}$ of $\phi$ meson as a function of the azimuthal angle at $T=0.146$ GeV for a fixed transverse momentum $p_T=1$ GeV at the mid rapidity $Y=0$ (solid red line) and more forward rapidity $Y=1$ (dashed blue line).  }
\label{azmphi}
\end{figure}

The result of $\rho_{00}$ as a function of the meson’s azimuthal angle for a fixed transverse momentum $p_T=1$ GeV at the mid rapidity $Y=0$ and more forward rapidity $Y=1$ is shown in Fig. \ref{azmphi}. At $\varphi=0$, the $\phi$ meson has tendency toward a transverse alignment $\rho_{00}<1/3$ for both $Y=0$ and $Y=1$, with a larger deviation from $1/3$ for $Y=1$. By increasing the azimuthal angle to $\varphi=\pi/2$, the $\rho_{00}$ positively deviates from the $1/3$ and at it is the maximum value. The negative deviation reappears when we reach to the $\varphi=\pi$. 

\begin{figure}
  \centering
  \includegraphics[width=0.55\linewidth]{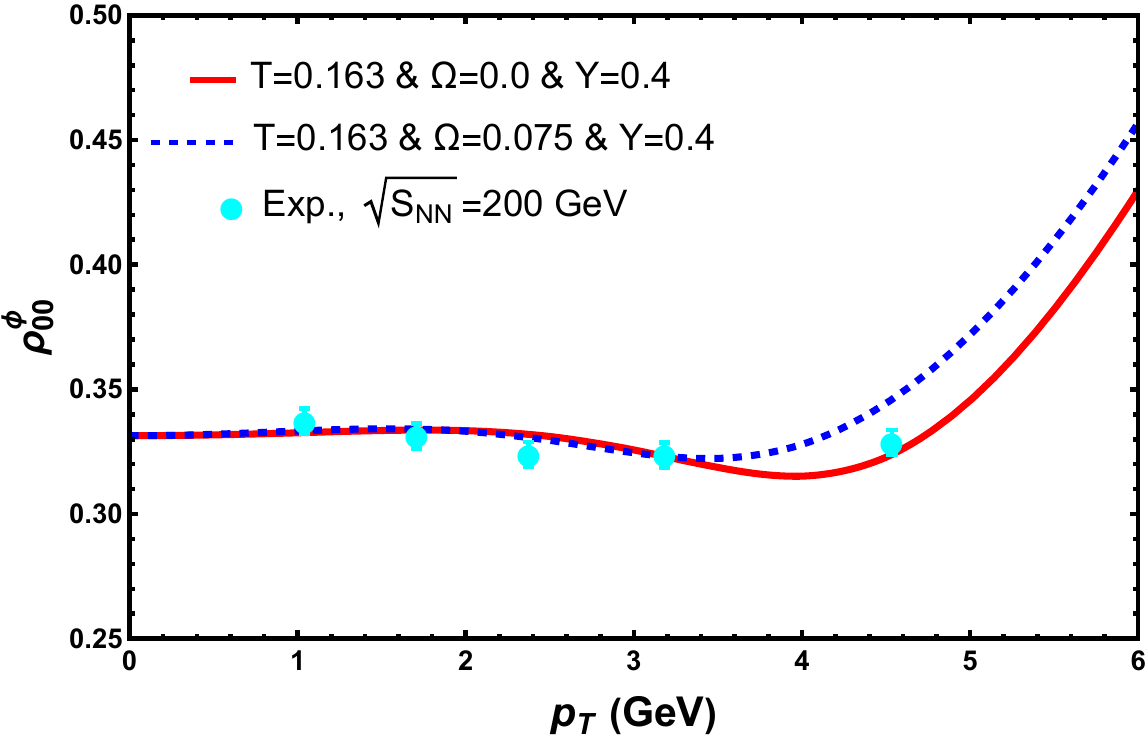} 
\caption{ The $\rho_{00}$ of $\phi$ meson as a function of $p_T$ at the rapidity $Y=0.4$ for a non-rotating thermal medium at $T=0.163$ Gev (solid red line) and a rotating thermal medium at $T=0.163$ GeV and $\Omega=0.075$ GeV (dashed blue line). The experimental data is taken from Ref. \cite{STAR:2022fan}. }
\label{spinphi}
\end{figure}

To examine the global spin alignment, we averaged the $\rho_{00}$ over the meson’s azimuthal angle $0 \le \varphi \le 2 \pi$, excluding the effect of the elliptic flow, and plotted the results as a function of transverse momentum in Fig. \ref{spinphi}. In this figure, we compare our results with experimental data at $\sqrt{s_{NN}}=200$ GeV and $|Y|<1$ \cite{STAR:2022fan}. We chose $\sqrt{s_{NN}}=200$ GeV from experimental data to compare with our results because the temperature of our thermal medium is high, corresponding to a higher center of mass energy \cite{Andronic:2017pug}.

We consider a specific case by fixing the rapidity at $Y=0.4$ and examine two scenarios: a non-rotating thermal medium at $T=0.163$ GeV and a rotating thermal medium at $T=0.163$ GeV and $\Omega=0.075$ GeV. In both scenarios,  $\rho_{00}$ is slightly greater than $1/3$ for $p_T<2$ GeV, reaching a minimum value around $p_T \approx 4$ GeV and $p_T \approx 3$ GeV, respectively. Additionally, at high $p_T>3$ GeV, rotation enhances the spin alignment of the $\phi$ meson.

\begin{figure}
  \centering
  \includegraphics[width=0.48\linewidth]{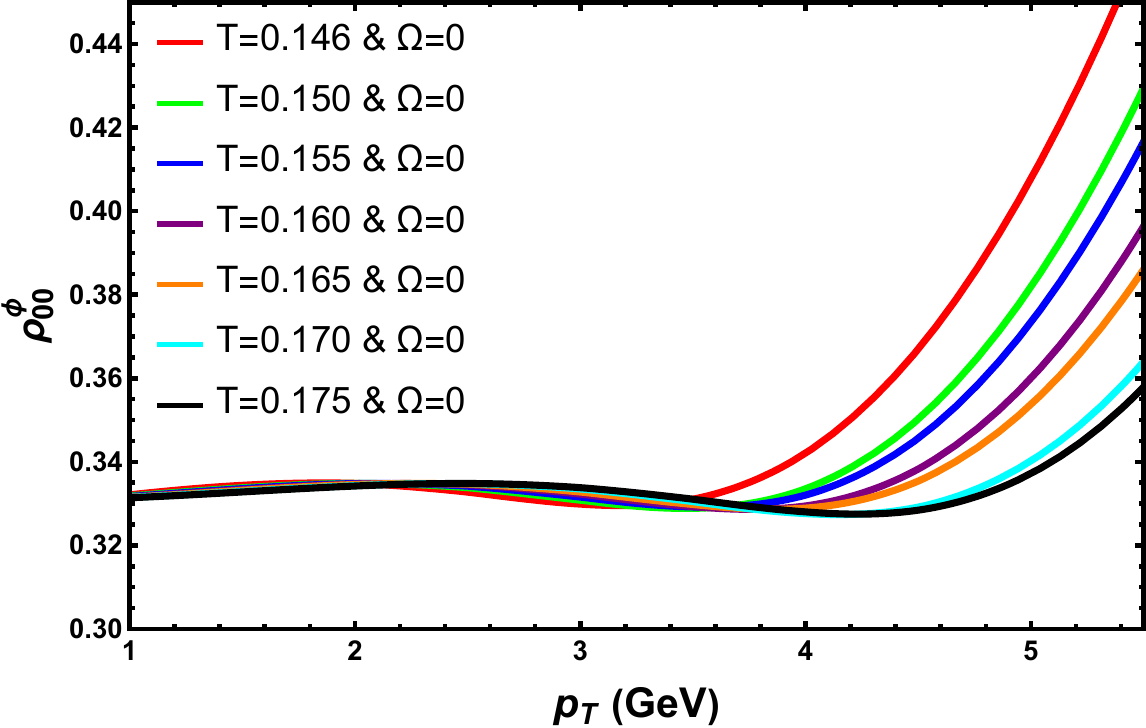} 
  \includegraphics[width=0.50\linewidth]{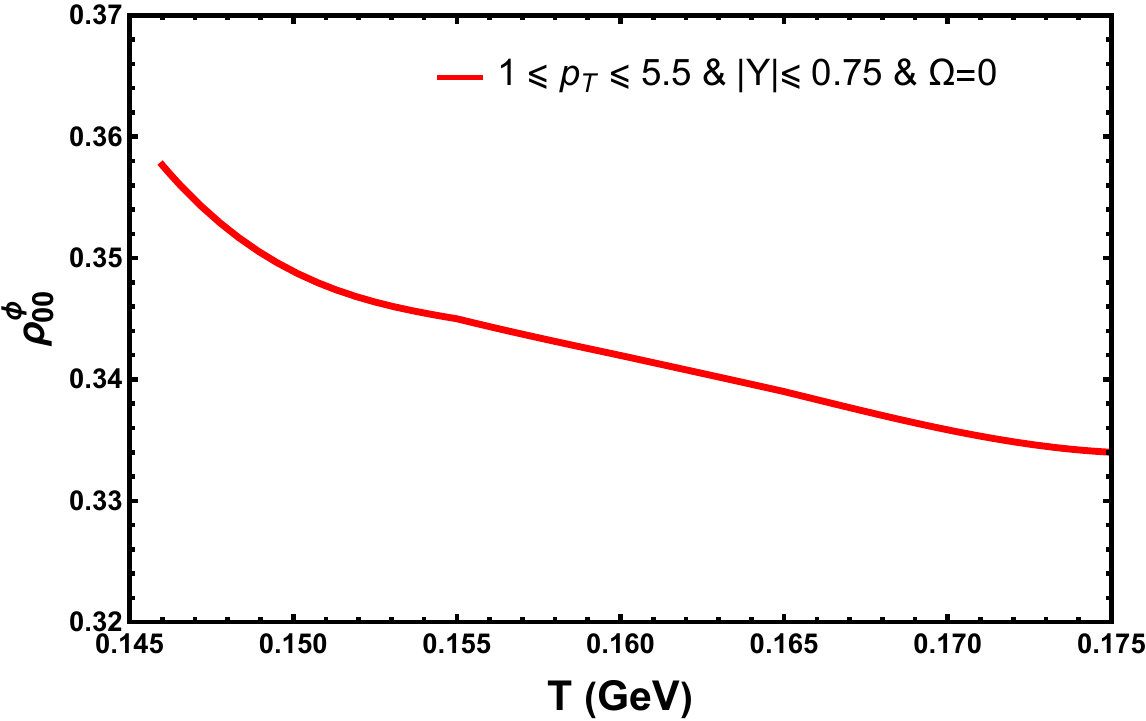}   
\caption{Left panel: The effect of temperature in a non-rotating background on the $\rho_{00}$ as a function of $p_T$ at the rapidity window $|Y| \le 0.75$ for $\phi$ meson. Right panel: The averaged $\rho_{00}$ over the transverse momentum region $ 1 \le p_T  \le 5.5$ GeV as a function of temperature.  }
\label{rho00phiT}
\end{figure}

To explicitly analyze the effect of temperature in a non-rotating background on the alignment of the $\phi$ meson, we computed the $\rho_{00}$ of the $\phi$ meson as a function of transverse momentum for a range of temperatures from $T = 0.146$ GeV to $T = 0.175$ GeV, as shown in Fig. \ref{rho00phiT}. The rapidity window was chosen as $|Y| \le 0.75$. For all temperatures, at low transverse momentum ($p_T \le 3.5$ GeV), the deviation of $\rho_{00}$ from $1/3$ is not significant and is almost independent of temperature, as illustrated in the left panel of Fig. \ref{rho00phiT}. In contrast, the temperature effect becomes more pronounced at high $p_T$. As the temperature increases, the alignment is suppressed. The right panel of Fig. \ref{rho00phiT} shows that the value of $\rho_{00}$ in the transverse momentum range of $1 \le p_T \le 5.5$ GeV consistently decreases with increasing temperature. This behavior is qualitatively similar to the experimental results of $\rho_{00}$ as a function of $\sqrt{s_{NN}}$ \cite{STAR:2022fan}.

\begin{figure}
  \centering
  \includegraphics[width=0.49\linewidth]{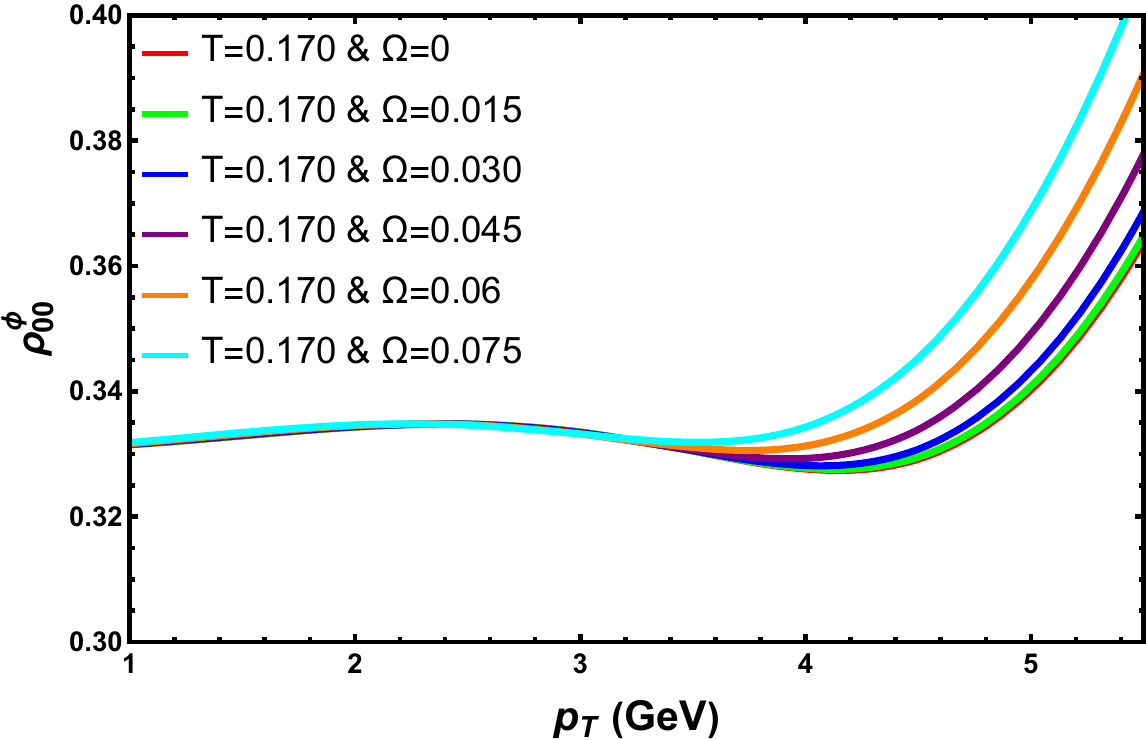} 
  \includegraphics[width=0.50\linewidth]{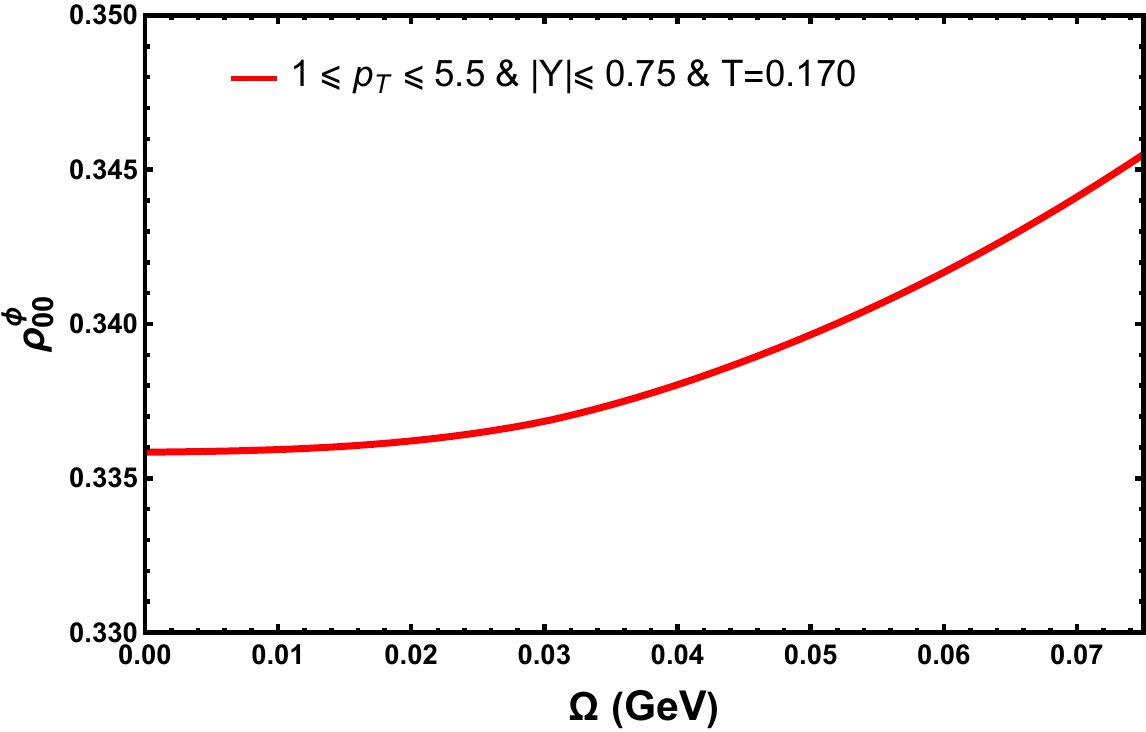}   
\caption{Left panel: The effect of rotation in a thermal equilibrium medium at $T=0.170$ GeV on the $\rho_{00}$ as a function of $p_T$ at the rapidity window $|Y| \le 0.75$ for $\phi$ meson. Right panel: The averaged $\rho_{00}$ over the transverse momentum region $ 1 \le p_T  \le 5.5$ GeV as a function of angular velocity.  }
\label{rho00phiomega}
\end{figure}

Additionally, we investigate how rotation affects the global spin alignment for a thermal medium at a fixed temperature of $T = 0.170$ GeV. In the left panel of Fig. \ref{rho00phiomega}, it can be seen that for low transverse momentum ($p_T \ge 3$ GeV), the value of $\rho_{00}$ is independent of the angular velocity. However, beyond a certain momentum threshold, the effect of rotation becomes evident and enhances the alignment. Overall, the value of $\rho_{00}$ in the range of $1 \le p_T \le 5.5$ GeV increases with increasing angular velocity. This can be understood in terms of transferring a part of the angular momentum of the background to the vector meson through spin-orbit couplings \cite{Liang:2004ph,Liang:2004xn}.

\begin{figure}
  \centering
  \includegraphics[width=0.55\linewidth]{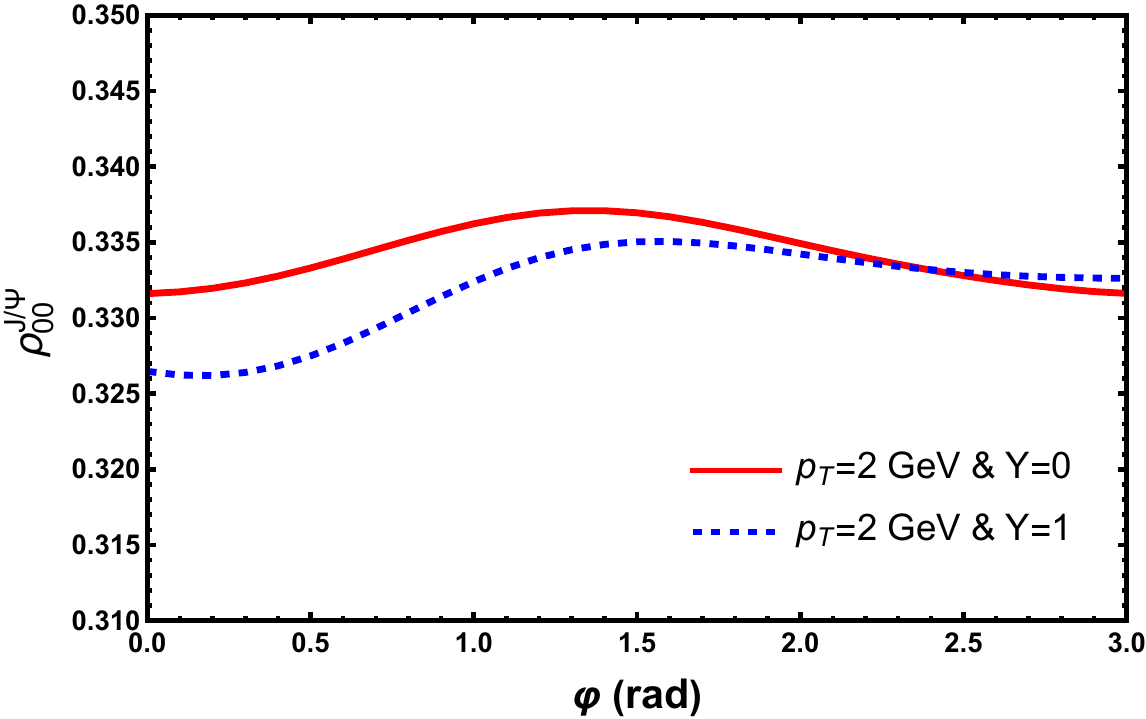} 
\caption{The $\rho_{00}$ of $J/\Psi$ meson as a function of the azimuthal angle at $T=0.146$ GeV for a fixed transverse momentum $p_T=2$ GeV at the mid rapidity $Y=0$ (solid red line) and more forward rapidity $Y=1$ (dashed blue line).  }
\label{azmjpsi}
\end{figure}

\begin{figure}
  \centering
  \includegraphics[width=0.55\linewidth]{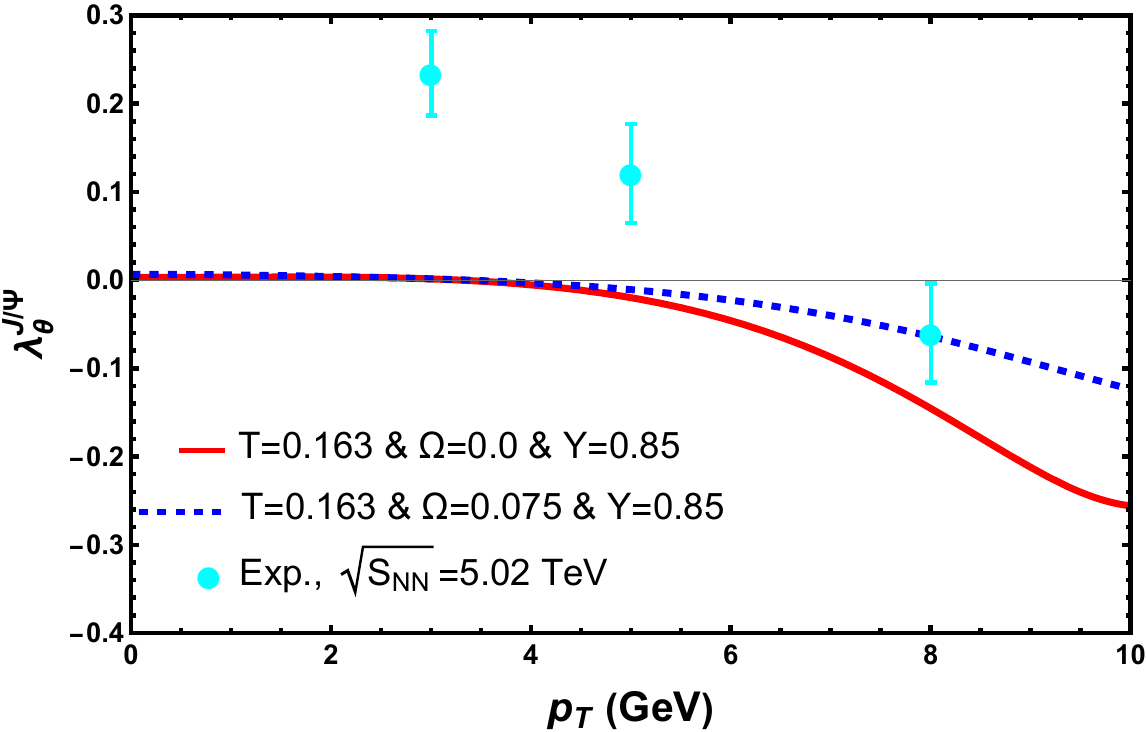}   
\caption{ The $\lambda_{\theta}$ of $J/\Psi$ meson as a function of $p_T$ at the rapidity $Y=0.85$ for a non-rotating thermal medium at $T=0.163$ Gev (solid red line) and a rotating thermal medium at $T=0.163$ GeV and $\Omega=0.075$ GeV (dashed blue line). The experimental data is taken from Ref. \cite{ALICE:2022dyy}.  }
\label{spinjpsi}
\end{figure}

Analogous to the investigation of the $\phi$ meson, we explore the global spin alignment of the $J/\Psi$. The azimuthal dependence of the $\rho_{00}$ for the $J/\Psi$ at rapidity values $Y=0$ and $Y=1$, specifically for a transverse momentum of $p_T = 2$ GeV, is illustrated in Fig. \ref{azmjpsi}. At $\varphi=0$, the value of $\rho_{00}$ stays below $1/3$, and only overpass $1/3$ for $Y=0$ at $\varphi=\pi/2$. This observed behavior aligns with experimental data, which indicates that for forward rapidity at lower $p_T$ values, the $\rho_{00}$ consistently remains less than $1/3$ \cite{ALICE:2022dyy}. Since the experimental collaboration has measured $\lambda_{\theta}$ for $J/\Psi$, we also switch to $\lambda_{\theta}$ to compare with it. After averaging over the azimuthal angle, we compare our results for the $\lambda_{\theta}$ as a function of $p_T$ at the rapidity $Y=0.85$ with the experimental data \cite{ALICE:2022dyy} as shown in Fig. \ref{spinjpsi}. In both scenarios, the examined results, from a non-rotating thermal medium at $T=0.163$ GeV and from a rotating thermal medium at $T=0.163$ GeV with $\Omega=0.075$ GeV, qualitatively agree with the experimental data for only high $p_T$. This is not surprising because of the fact that the $\lambda_{\theta}$ has been measured in the forward rapidity region $2.5 \le Y \le 4$, which is different from our chosen setup. This would be interesting if the experimental collaboration can measure the $\lambda_{\theta}$ in the mid-rapidity region.

\begin{figure}
  \centering
  \includegraphics[width=0.47\linewidth]{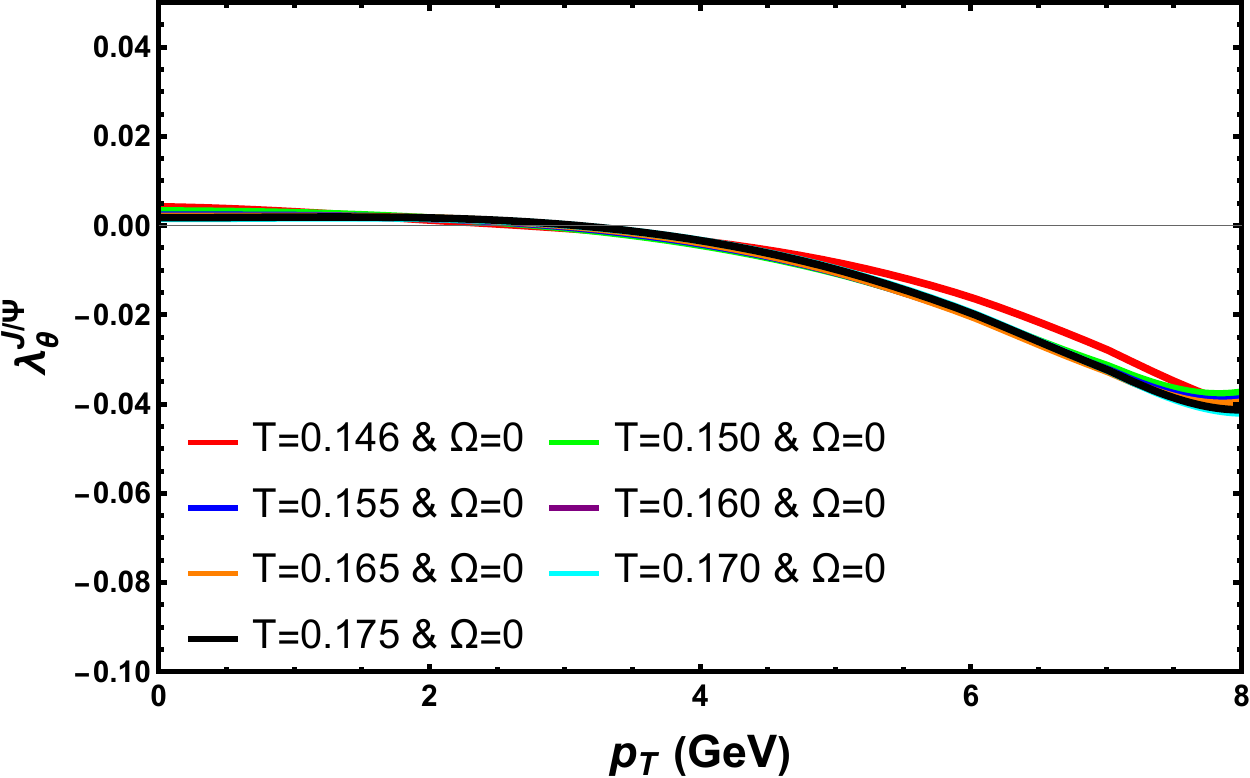} 
  \includegraphics[width=0.51\linewidth]{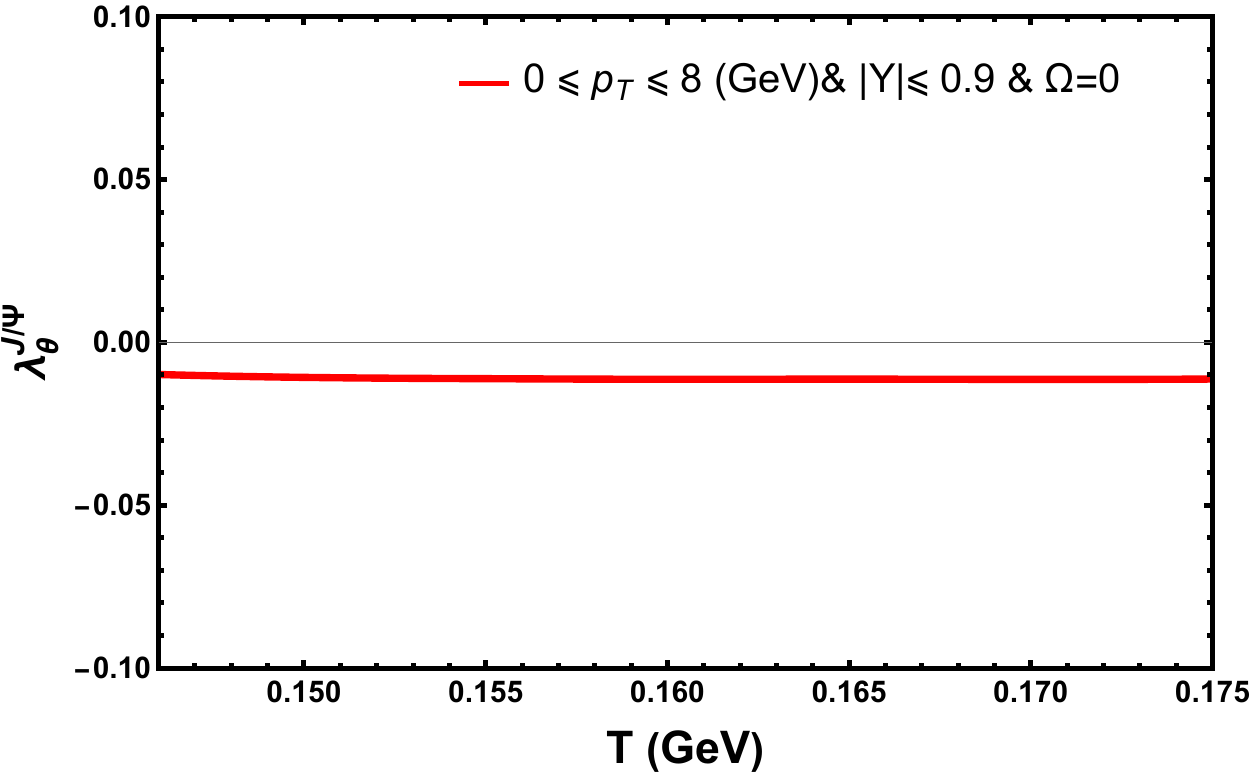}   
\caption{Left panel: The effect of temperature in a non-rotating background on the $\lambda_{\theta}$ of $J/\Psi$ as a function of $p_T$ at the rapidity window $|Y| \ge 0.9$. Right panel: The averaged $\lambda_{\theta}$ over the transverse momentum region $ 0 \le p_T  \le 8$ GeV as a function of temperature.  }
\label{rho00jpsiT}
\end{figure} 

\begin{figure}
  \centering
  \includegraphics[width=0.48\linewidth]{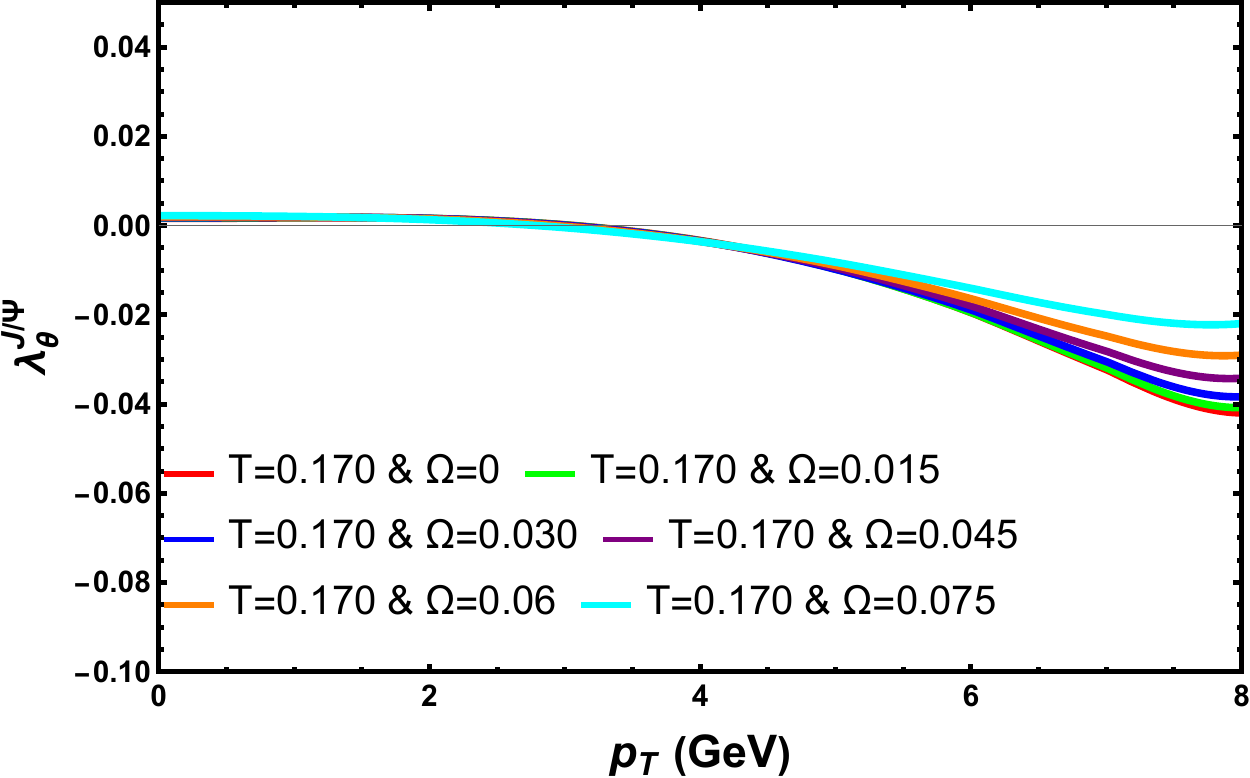} 
  \includegraphics[width=0.50\linewidth]{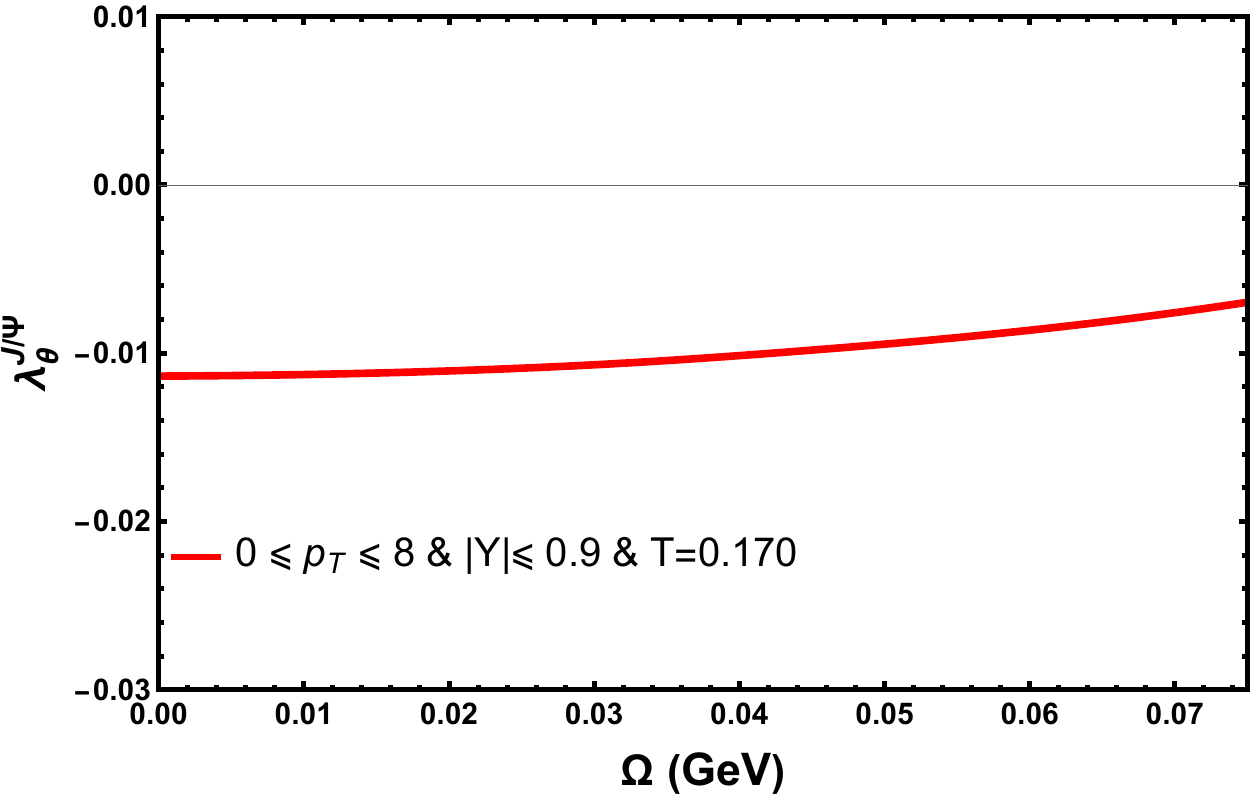}   
\caption{Left panel: The effect of rotation in a thermal equilibrium medium at $T=0.170$ GeV on the $\lambda_{\theta}$ of $J/\Psi$ as a function of $p_T$ at the rapidity window $|Y| \ge 0.9$. Right panel: The averaged $\lambda_{\theta}$ over the transverse momentum region $ 0 \le p_T  \le 8$ GeV as a function of angular velocity.  }
\label{rho00jpsiomega}
\end{figure} 

Considering the effect of the thermal fluctuation on the $\lambda_{\theta}$ of the $J/\Psi$, we show the results of the average $\lambda_{\theta}$ over the full azimuthal angle range and rapidity ($|Y| \le 0.9$) as a function of $p_T$ at different temperature in Fig. \ref{rho00jpsiT}. From Fig. \ref{rho00jpsiT}, one can see that the $\lambda_{\theta}$ is not sensitive to the temperature variations in the range $T=0.146-0.175$ GeV, and stays below zero for all temperatures. This behavior is consistent with the fact that due to the large mass of the charm quark, the heavy quarkonium state $J/\Psi$ is less susceptible to the medium effect. Similarly, the $\lambda_{\theta}$ as a function of $p_T$ for different angular velocity ranging from $\Omega=0.0-0.075$ GeV is shown in Fig. \ref{rho00jpsiomega}. The $\lambda_{\theta}$ is insensitive to the variation of $\Omega$ up to $p_t=5$ GeV, at high transverse momentum, $\lambda_{\theta}$ shows a slight suppression for increasing $\Omega$. The averaged $\lambda_{\theta}$ over the $p_T$ range ($0 \le p_t \le 8$) as a function of $\Omega$ illustrate the behavior more clearly, see the left panel of Fig. \ref{rho00jpsiomega}. Similar behavior for the spin alignment of $J/\Psi$ has been reported in the magnetic field background \cite{Zhao:2024ipr}. This nontrivial feature of the $J/\Psi$ meson needs more investigation, which is beyond the scope of the current work.

\begin{figure}
  \centering
  \includegraphics[width=0.48\linewidth]{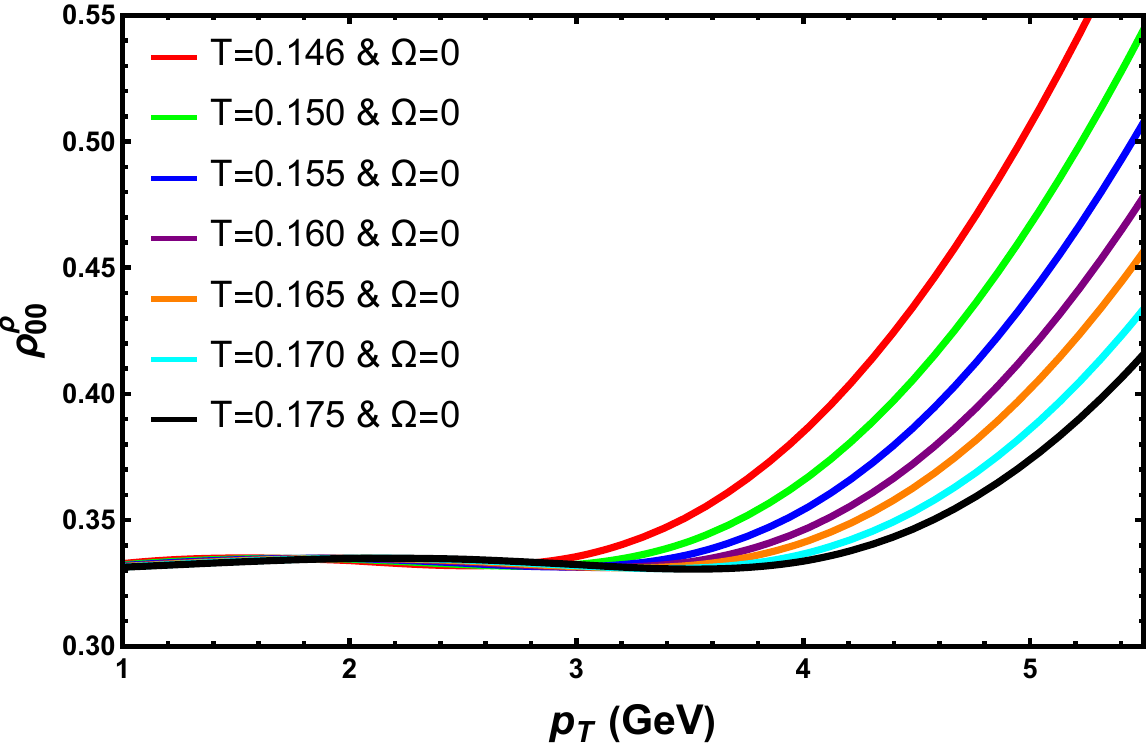} 
  \includegraphics[width=0.50\linewidth]{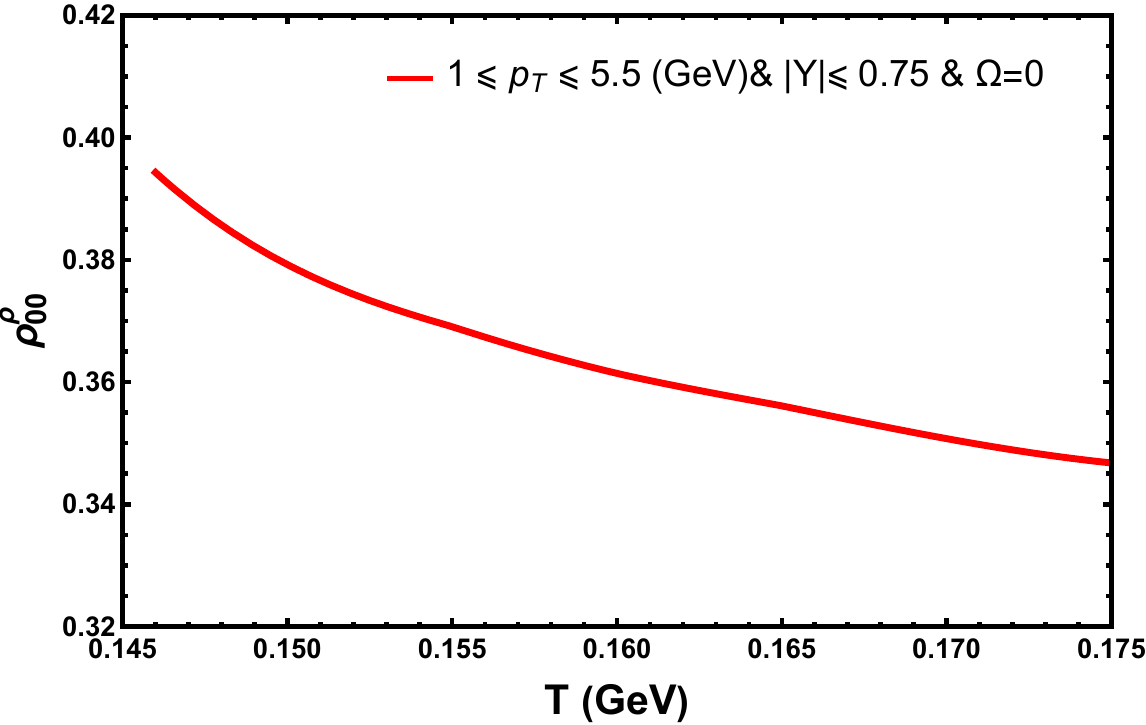}   
\caption{Left panel: The effect of temperature in a non-rotating background on the $\rho_{00}$ of $\rho$ meson as a function of $p_T$ at the rapidity window $|Y| \le 0.75$. Right panel: The averaged $\rho_{00}$ over the transverse momentum region $ 1 \le p_T  \le 5.5$ GeV as a function of temperature.  }
\label{rho00rhoT}
\end{figure} 

\begin{figure}
  \centering
  \includegraphics[width=0.49\linewidth]{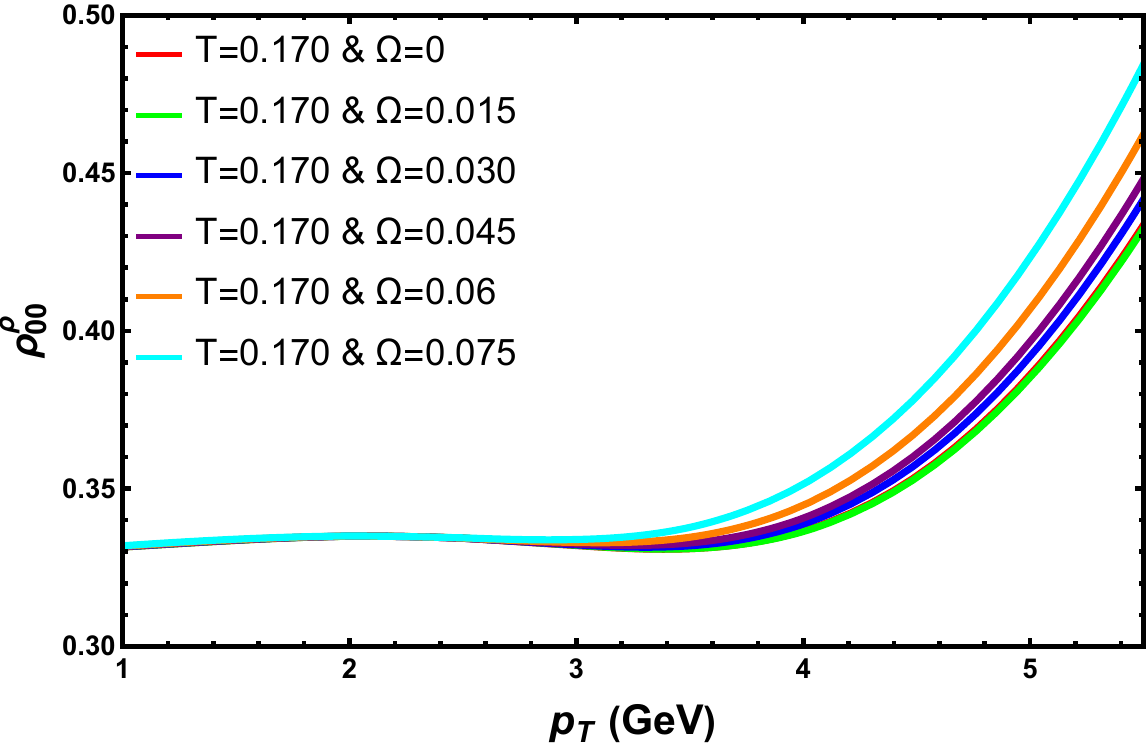} 
  \includegraphics[width=0.50\linewidth]{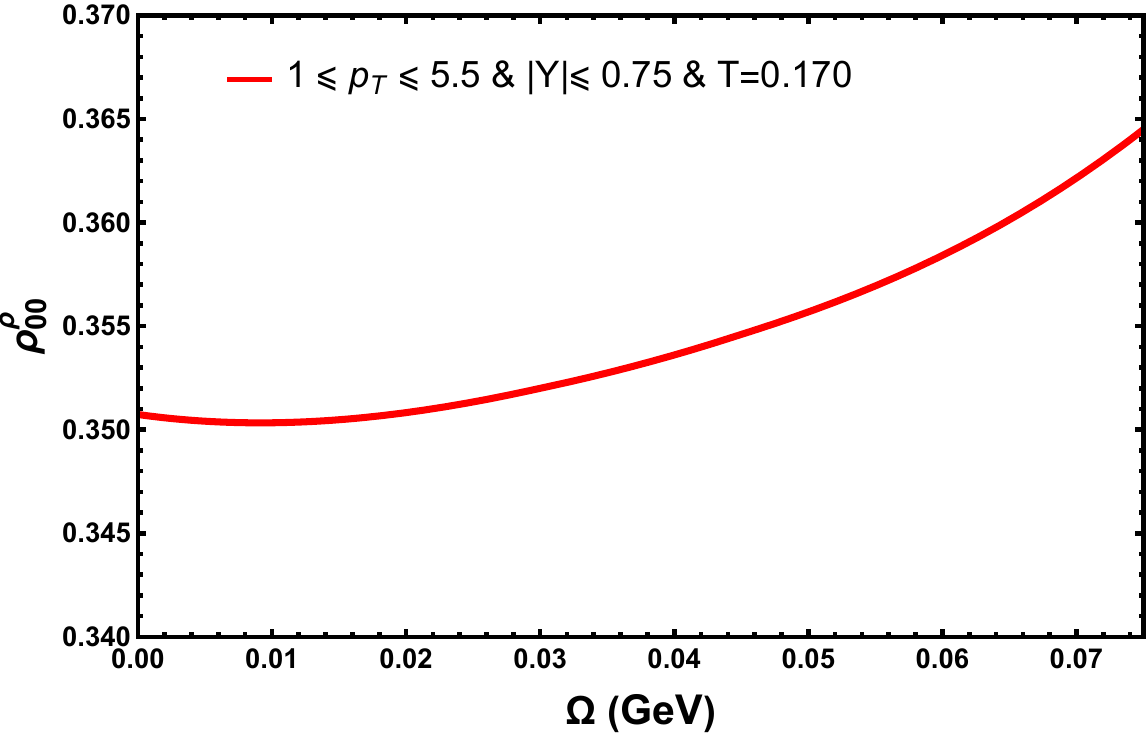}   
\caption{Left panel: The effect of rotation in a thermal equilibrium medium at $T=0.170$ GeV on the $\rho_{00}$ of $\rho$ meson as a function of $p_T$ at the rapidity window $|Y| \le 0.75$. Right panel: The averaged $\rho_{00}$ over the transverse momentum region $ 1 \le p_T  \le 5.5$ GeV as a function of angular velocity.  }
\label{rho00rhoomega}
\end{figure}  

Although there is currently no measured signal for the global spin alignment of the $\rho$ meson, we can present a benchmark result for the zeroth component of the spin density matrix, $\rho_{00}$, through the dilepton decay channel of the $\rho$ meson. The dependence of $\rho_{00}$ on transverse momentum ($p_T$) is illustrated in Fig. \ref{rho00rhoT}. In this analysis, we have chosen a rapidity window of $|Y| \le 0.75$ across different temperatures. Similar to the scenario observed with the $\phi$ meson, the $\rho$ meson shows no alignment for low transverse momenta ($p_T \le 3$ GeV); however, beyond this critical point, the alignment grows exponentially with increasing $p_T$. From the left panel of Fig. \ref{rho00rhoT}, it is evident that thermal fluctuations dampen the growth of $\rho_{00}$. This behavior is further illustrated in the right panel, where the average $\rho_{00}$ over the transverse momentum range of $1 \le p_T \le 5.5$ GeV displays a smooth decrease with rising temperature. When considering the influence of rotation at a fixed temperature, we observe an enhancement in global alignment, as depicted in Fig. \ref{rho00rhoomega}. 

By comparing the alignment of the $\rho$ meson with that of the $\phi$ meson, we can see that while their qualitative behaviors are similar, the medium effects on the $\rho$ meson are more pronounced. We anticipate that, in the near future, the spin alignment of the $\rho$ meson will be measured to validate our predicted results.

It is important to keep in mind that the difference between the spin alignment of the vector mesons in holographic QCD is due to their different spectral functions. This might be a clue that the spin alignment of the mesons is a nonperturbative property in the strong coupling medium.

\section{Conclusions} 
\label{sectionIIII}

In this paper, we have investigated the behaviour of the vector mesons ($\rho$, $\phi$, and $J/\Psi$) in a non-rotating thermal medium and a rotating thermal medium by analyzing their spectral functions and spin alignment properties. To utilize the effects of the rotation, we introduced anisotropic backgrounds that can be obtained by solving the EMD action. The effect of the rotation or the angular velocity $\Omega$ is introduced into the anisotropic backgrounds through the $U(1)$ gauge field and the induced gluon polarization is captured by the $\Omega$-dependent dilaton field. In order to clarify the effect of temperature and rotation on the vector mesons, we have employed the soft-wall holographic QCD model with four flavors.

Considering the fact that we have a pure gluonic background, the order of the deconfinement/confinement transition is the first order with zero chemical potential. By adding the chemical potential to the non-rotating background, the order of the phase transition changes from the first order to crossover. The point where the order of the phase transition changes marks the CEP, which is located at ($T_{CEP}, \mu_{CEP}) = (0.1128, 0.45)$ GeV. Moreover, adding the rotation to the medium shifts the CEP to a higher value. Keeping the chemical potential at zero and adding the rotation to the background increases the deconfinement/confinement transition temperature, which is consistent with the LQCD data. Additionally, the behavior of the critical temperature $T_c$ with the imaginary angular velocity $\Omega_I$ can be fitted with a simple quadratic function.

The spectral functions of the $\phi$ and $\rho$ mesons exhibit broad peaks at lower temperatures, indicating their presence in the medium. These peaks disappear at higher temperatures, signaling the melting of the mesons. However, in a rotating medium, the dissociation temperature increases, as rotation delays the melting process. For the $J/\Psi$ meson, a prominent peak in the spectral function suggests a quasi-particle state. Unlike the $\phi$ and $\rho$ mesons, the $J/\Psi$ meson is more resilient to thermal effects due to its heavy charm quark content.

Finally, we study the global spin alignment of $\phi$, $J/\Psi$, and $\rho$ mesons in the event plane frame. For $\phi$ meson, the spin alignment observable $\rho_{00}$ exhibits a strong dependence on the azimuthal angle $\varphi$, with deviation from $1/3$ indicating transverse alignment at $\varphi=0$ and $\varphi=\pi$, and positive deviation at $\varphi=\pi/2$. The averaged $\rho_{00}$ over the full range of the azimuthal angle at low $p_T$ shows a slight dependence on the temperature. Moving to the high $p_T$, the temperature effect becomes significant and reduces the $\phi$ meson alignment. This global feature is consistent with the dampening of the experimental measurement of $\rho_{00}$ as a function of the center of mass energy. Similarly, the rotation effect is emergent at high $p_T$, while $\rho_{00}$ enhanced in the rotating medium, which can be understood as the transfer of the angular momentum of the medium to the $\phi$ meson through spin-orbit coupling.

For $J/\Psi$ meson, the spin alignment behavior differs due to its heavy quark nature. The $\lambda_{\theta}$ remains largely insensitive to temperature variations and angular velocity up to $p_T=5$ GeV, beyond which rotation slightly suppresses the alignment. This insensitivity is attributed to the large mass of the charm quark, making $J/\Psi$ less susceptible to medium effects. However, the observed suppression at high $p_T$ warrants further investigation to determine if it is a general feature of quarkonia. The difference between the spin alignment behaviour for $\phi$ and $J/\Psi$ mesons appears because of the difference between their spectral functions. 

Lastly, the spin alignment of $\rho$ meson, though not yet experimentally measured, shows similar qualitative behavior to the $\phi$ meson, with no alignment at low $p_T$ and exponential growth in alignment at high $p_T$. Thermal fluctuations dampen this growth, while rotation enhances it. The medium effects on the $\rho$ meson are more pronounced compared to the $\phi$ meson, highlighting the need for future experimental measurements to validate these predictions. Overall, these results provide a comprehensive understanding of spin alignment in vector mesons, emphasizing the roles of temperature, rotation, and medium effects in heavy ion collisions.

\section*{Acknowledgments}

This work is supported in part by the National Natural Science Foundation of China (NSFC) Grant Nos: 12235016, 12221005, 12305136 and the Strategic Priority Research Program of Chinese Academy of Sciences under Grant No XDB34030000, the start-up funding from University of Chinese Academy of Sciences(UCAS), the start-up funding of Hangzhou Normal University under Grant No. 4245C50223204075, and the Fundamental Research Funds for the Central Universities. H. A. A. acknowledges the "Alliance of International Science Organization (ANSO) Scholarship For Young Talents" for providing financial support for the Ph.D. study.

\appendix
\section{ Spectral function in spin space }
\label{sfss}

The spectral function in the spin space of the dimuon is given by Eq. \eqref{spinspectral}. Here, we provide the details of the components which can be accessible for any particular frame and spin state.

\begin{equation}
 \begin{aligned}
\tilde{ \varrho}_{\lambda \lambda^{\prime}}(p)  =&  \eta_{\mu \alpha} \eta_{\nu \beta} v^{* \alpha}(\lambda, p) v^\beta\left(\lambda^{\prime}, p\right) \varrho^{\mu \nu}(p) \\
                                        =& - v^{* 0}(\lambda) v^0\left(\lambda^{\prime}\right) \operatorname{Im} D^{0 0}+v^{* 0}(\lambda) v^j\left(\lambda^{\prime}\right) \operatorname{Im} D^{0 j}+v^{* i}(\lambda) v^0\left(\lambda^{\prime}\right) \operatorname{Im} D^{i 0}-v^{* i}(\lambda) v^j\left(\lambda^{\prime}\right) \operatorname{Im} D^{i j} \\
                                       = &  \frac{1}{g_{5}^{2}} \frac{e^{-\phi(z)} f(z) h(\phi)}{z} \operatorname{Im}\left[  \right.\\
                             &\left. \left( v^{* 0}(\lambda) v^0\left(\lambda^{\prime}\right) p_{x_1}^{2}- (v^{* 0}(\lambda) v^{x_1}\left(\lambda^{\prime}\right) + v^{* x_1}(\lambda) v^0\left(\lambda^{\prime}\right) )\omega p_{x_1}  +v^{* x_1}(\lambda) v^{x_1}\left(\lambda^{\prime}\right) \omega^{2}  \right)  (1 - \frac{p_{x_1}^2}{p^2}) \mathcal{E}_{x_1}(z) \partial_z \mathcal{E}_{x_1}(z)   \right.\\
                             &\left. + \left( v^{* 0}(\lambda) v^0\left(\lambda^{\prime}\right) p_{x_2}^{2}- (v^{* 0}(\lambda) v^{x_2}\left(\lambda^{\prime}\right) + v^{* x_2}(\lambda) v^0\left(\lambda^{\prime}\right) )\omega p_{x_2}  +v^{* x_2}(\lambda) v^{x_2}\left(\lambda^{\prime}\right) \omega^{2}  \right)  (1 - \frac{p_{x_2}^2}{p^2}) \mathcal{E}_{x_2}(z) \partial_z \mathcal{E}_{x_2}(z)   \right.\\
                             &\left. +  \left( v^{* 0}(\lambda) v^0\left(\lambda^{\prime}\right) p_{x_3}^{2}- (v^{* 0}(\lambda) v^{x_3}\left(\lambda^{\prime}\right) + v^{* x_3}(\lambda) v^0\left(\lambda^{\prime}\right) )\omega p_{x_3}  +v^{* x_3}(\lambda) v^{x_3}\left(\lambda^{\prime}\right) \omega^{2}  \right)  (1 - \frac{p_{x_3}^2}{p^2}) \mathcal{E}_{x_3}(z) \partial_z \mathcal{E}_{x_3}(z) \right.\\
                             &\left. - p_{x_1} p_{x_2} \left( v^{* 0}(\lambda) v^0\left(\lambda^{\prime}\right) \frac{p_{x_1} p_{x_2}}{p^2} - (v^{* 0}(\lambda) v^{x_2}\left(\lambda^{\prime}\right) + v^{* x_2}(\lambda) v^0\left(\lambda^{\prime}\right)) \frac{p_{x_1} \omega}{p^2}  +v^{* x_1}(\lambda) v^{x_2}\left(\lambda^{\prime}\right) \frac{\omega^{2}}{p^2}  \right) \mathcal{E}_{x_1}(z) \partial_z \mathcal{E}_{x_2}(z) \right.\\
                             &\left. - p_{x_1} p_{x_2} \left( v^{* 0}(\lambda) v^0\left(\lambda^{\prime}\right) \frac{p_{x_1} p_{x_2}}{p^2} - (v^{* 0}(\lambda) v^{x_1}\left(\lambda^{\prime}\right)  + v^{* x_1}(\lambda) v^0\left(\lambda^{\prime}\right)) \frac{p_{x_2} \omega}{p^2}  +v^{* x_1}(\lambda) v^{x_2}\left(\lambda^{\prime}\right) \frac{\omega^{2}}{p^2}  \right) \mathcal{E}_{x_2}(z) \partial_z \mathcal{E}_{x_1}(z)  \right.\\
                             &\left. - p_{x_1} p_{x_3} \left( v^{* 0}(\lambda) v^0\left(\lambda^{\prime}\right) \frac{p_{x_1} p_{x_3}}{p^2} - (v^{* 0}(\lambda) v^{x_3}\left(\lambda^{\prime}\right) + v^{* x_3}(\lambda) v^0\left(\lambda^{\prime}\right)) \frac{p_{x_1} \omega}{p^2}  +v^{* x_1}(\lambda) v^{x_3}\left(\lambda^{\prime}\right) \frac{\omega^{2}}{p^2}  \right) \mathcal{E}_{x_1}(z) \partial_z \mathcal{E}_{x_3}(z)   \right.\\
                             &\left. - p_{x_1} p_{x_3} \left( v^{* 0}(\lambda) v^0\left(\lambda^{\prime}\right) \frac{p_{x_1} p_{x_3}}{p^2} - (v^{* 0}(\lambda) v^{x_1}\left(\lambda^{\prime}\right)  + v^{* x_1}(\lambda) v^0\left(\lambda^{\prime}\right)) \frac{p_{x_3} \omega}{p^2}  +v^{* x_1}(\lambda) v^{x_3}\left(\lambda^{\prime}\right) \frac{\omega^{2}}{p^2}  \right) \mathcal{E}_{x_3}(z) \partial_z \mathcal{E}_{x_1}(z) \right.\\
                             &\left. - p_{x_2} p_{x_3} \left( v^{* 0}(\lambda) v^0\left(\lambda^{\prime}\right) \frac{p_{x_2} p_{x_3}}{p^2} - (v^{* 0}(\lambda) v^{x_3}\left(\lambda^{\prime}\right) + v^{* x_3}(\lambda) v^0\left(\lambda^{\prime}\right)) \frac{p_{x_2} \omega}{p^2}  +v^{* x_2}(\lambda) v^{x_3}\left(\lambda^{\prime}\right) \frac{\omega^{2}}{p^2}  \right) \mathcal{E}_{x_2}(z) \partial_z \mathcal{E}_{x_3}(z)   \right.\\
                             &\left. - p_{x_2} p_{x_3} \left( v^{* 0}(\lambda) v^0\left(\lambda^{\prime}\right) \frac{p_{x_2} p_{x_3}}{p^2} - (v^{* 0}(\lambda) v^{x_2}\left(\lambda^{\prime}\right)  + v^{* x_2}(\lambda) v^0\left(\lambda^{\prime}\right)) \frac{p_{x_3} \omega}{p^2}  +v^{* x_2}(\lambda) v^{x_3}\left(\lambda^{\prime}\right) \frac{\omega^{2}}{p^2}  \right) \mathcal{E}_{x_3}(z) \partial_z \mathcal{E}_{x_2}(z) \left. \right] \right|_{\epsilon}. 
 \end{aligned}
\end{equation}

The components of the spin polarization vector $ v^{\mu}(\lambda, p)$ depend on the framework choice. In the current work, we consider $x_2$ as the direction of the spin quantization of the vector meson (which in heavy ion collision is known as the event-plane frame), then $\epsilon_{\lambda}$ is given by 

\begin{equation}
\begin{aligned}
\boldsymbol{\epsilon}_0 & =(0,1,0), \\
\boldsymbol{\epsilon}_{+1} & =-\frac{1}{\sqrt{2}}(1, 0, i), \\
\boldsymbol{\epsilon}_{-1} & =\frac{1}{\sqrt{2}}(1,0, -i),
\end{aligned}
\end{equation}
thereafter the covariant spin polarization vector is obtained as 

\begin{equation}
\begin{aligned}
& v^\mu(0, p)=\left(\frac{p_{x_2}}{M}, \frac{p_{x_2}}{M(\omega+M)} p_{x_1} ,1+ \frac{p_{x_2}}{M(\omega+M)} p_{x_2} , \frac{p_{x_2}}{M(\omega+M)} p_{x_3}\right) \\
& v^\mu(+1, p)=\left(\frac{-1}{\sqrt{2}}\frac{(p_{x_1} + i p_{x_3})}{M}, \frac{-1}{\sqrt{2}}-\frac{(p_{x_1} + i p_{x_3})}{\sqrt{2}M(\omega+M)}  p_{x_1}, -\frac{(p_{x_1} + i p_{x_3})}{\sqrt{2} M(\omega+M)}  p_{x_2} , \frac{-i}{\sqrt{2}}-\frac{(p_{x_1} + i p_{x_3})}{\sqrt{2} M(\omega+M)}  p_{x_3}  \right) \\
& v^\mu(-1, p)=\left(\frac{1}{\sqrt{2}}\frac{(p_{x_1} - i p_{x_3})}{M}, \frac{1}{\sqrt{2}}+\frac{(p_{x_1} - i p_{x_3})}{\sqrt{2}M(\omega+M)}  p_{x_1}, \frac{(p_{x_1} - i p_{x_3})}{\sqrt{2} M(\omega+M)}  p_{x_2} , \frac{i}{\sqrt{2}}+\frac{(p_{x_1} - i p_{x_3})}{\sqrt{2} M(\omega+M)}  p_{x_3}  \right)  .
\end{aligned}
\end{equation}


%

\addcontentsline{toc}{section}{References}

\end{document}